\documentclass{aa}  
\usepackage{graphicx,color,subfigure}
\usepackage{txfonts}
\usepackage{verbatim}
\graphicspath{{./figures/}}
\usepackage{natbib,twoopt}
\usepackage[breaklinks=true]{hyperref} 

\hypersetup{
	colorlinks=true,   
	urlcolor=blue,     
	linkcolor=blue     
}
\bibpunct{(}{)}{;}{a}{}{,} 

\makeatletter
\newcommand{\bibnote}[2]{\global\@namedef{#1note}{#2}}
\newcommand{\biblink}[2]{\global\@namedef{#1link}{#2}}
\makeatother
\makeatletter
\protected\def\stonyslink{%
	\def\hyper@linkstart##1##2{}\let\hyper@linkend\@empty}
\newcommandtwoopt{\citeads}[3][][]{%
	\href{http://adsabs.harvard.edu/abs/#3}%
	{\stonyslink \citealp[#1][#2]{#3}}
	\biblink{#3}{\href{http://adsabs.harvard.edu/abs/#3}{ADS}}}
\newcommandtwoopt{\citepads}[3][][]{%
	\href{http://adsabs.harvard.edu/abs/#3}%
	{\stonyslink \citep[#1][#2]{#3}}
	\biblink{#3}{\href{http://adsabs.harvard.edu/abs/#3}{ADS}}}
\newcommandtwoopt{\citetads}[3][][]{%
	\href{http://adsabs.harvard.edu/abs/#3}%
	{\stonyslink \citet[#1][#2]{#3}}
	\biblink{#3}{\href{http://adsabs.harvard.edu/abs/#3}{ADS}}}
\newcommandtwoopt{\citeyearads}[3][][]{%
	\href{http://adsabs.harvard.edu/abs/#3}%
	{\stonyslink \citeyear[#1][#2]{#3}}
	\biblink{#3}{\href{http://adsabs.harvard.edu/abs/#3}{ADS}}}
\makeatother

\begin{document}

\title{Secular Structure of 1:2 and 1:3 Resonances with Neptune}
\author{Hailiang Li\inst{1,2}
	\and
	Li-Yong Zhou\inst{1}\fnmsep\inst{2}
	}
\authorrunning{Li \& Zhou}
\offprints{L.-Y. Zhou, \email zhouly@nju.edu.cn}
\institute{School of Astronomy and Space Science, Nanjing University, 163 Xianlin Avenue, Nanjing 210046, China
	\and
	Key Laboratory of Modern Astronomy and Astrophysics in Ministry of Education, Nanjing University, China
	}
\date{}

\abstract{
The 1:N mean motion resonances with Neptune are of particular interest because they have two asymmetric resonance islands, where the distribution of trapped objects may bear important clues to the history of the Solar System. To explore the dynamics of these resonances and to investigate whether the imprints left by the early stage evolution can be preserved in the resonances, we conduct a comprehensive analyses on the 1:2 and 1:3 resonances. Adopt mainly the frequency analysis method, we calculate the proper frequencies of the motion of objects in the resonances, with which the secular mechanisms that influence the dynamics are determined. Use the spectral number as an indicator of orbital regularity, we construct dynamical maps on representative planes. By comparing the structures in the maps with the locations of the secular mechanisms,  we find that the  von-Zeipel-Lidov-Kozai mechanism and the $g=2s$ mechanism destabilize the influenced orbits and portray the overall structure of the 1:2 and 1:3 resonances. The secular resonance $2g-s=s_8$ opens a channel for objects to switch between the leading and trailing resonance islands, which can alter the population ratio between these two islands. The secondary resonances associated with the quasi 2:1 resonance between Uranus and Neptune are also detected, and they introduce more chaos to the motion. The fine dynamical structures of the 1:2 and 1:3 resonances revealed in this paper, combined with knowledge of resonant capture, provide a compelling explanation for the eccentricity distribution of observed Twotinos. And we anticipate a more complete understanding of the history of planetary migration in the Solar System can be achieved by comparing the results in this paper with the populations in the 1:N resonances in future when further observations bring us more objects.
}

\maketitle{}

\section{Introduction}
Trans-Neptunian objects (TNOs) are precious historical remains from the formation and evolution of our Solar System. Residing distantly from the Sun and major planets, they have the longest dynamical timescale and thus may contain the most primitive information about the origin of the Solar System. Among all TNOs, those resonant ones that are in mean motion resonances (MMRs) with Neptune are of particular interests because of their unique dynamical properties. The studies on their orbital stability and on the structure of resonant region provide valuable insights into the nature of TNOs.  So far, nearly one thousand of resonant TNOs have been discovered, accounting for 20\% population of all known TNOs. Plutinos, a subcategory of resonant TNOs who are in the 2:3 resonance like Pluto, make up about half of this population. In addition, dozens of resonances have been found to host small bodies, all the way up to hundreds of AU from the Sun\footnote{https://www.johnstonsarchive.net/astro/tnoslist.html}.

The Neptunian exterior resonances have been extensively studied in literature, with their resonance centres, widths, angular libration amplitudes, libration periods, resonance regions etc being carefully analysed \citepads[e.g.][]{Malhotra1996, Gallardo2006b, Saillenfest2016, Lan2019, Gallardo2019, Gallardo2020}. The 1:N resonance has received particular attention. Unlike other resonances for which the critical resonant angle librates symmetrically around $0^\circ$ or $180^\circ$, a 1:N resonance has two extra asymmetric resonance islands (around other than $0^\circ$ or $180^\circ$), making them more complex \citepads[see e.g.][]{Message1958, frangakis1973, Beauge1994, Malhotra1996, Kotoulas2005, Voyatzis2005a}. The asymmetric islands are known as the leading and trailing islands, depending on whether their resonance centre is less than or greater than $180^\circ$. The island is called `leading' (`trailing') because an object within it  is located ahead (behind) of Neptune in longitude when it reaches its perihelion, where it is most likely to be discovered. In addition to the symmetric libration around $0^\circ$ (or $180^\circ$) and asymmetric libration around the leading (or trailing) island, there is another resonance configuration that the motion wraps both asymmetric islands with a large libration amplitude. The trajectories of objects in such configuration are similar to the `horseshoe orbit' in the 1:1 resonance. Thus in this paper, we refer to this configuration as `horseshoe resonance' to distinguish it from the symmetric resonance that evolves around only one stable symmetric centre.

Among all 1:N resonances, the 1:2 resonance has the lowest order, nearest distance, and the largest observed population of objects (known as Twotinos), thus is particularly noteworthy. The planetary migration of Neptune can break the symmetry between the leading and trailing islands of the 1:2 resonance \citepads[see e.g.][]{Chiang2002,MurrayClay2005,Li2023}, and may result in a difference in the population of objects between the two asymmetric islands. However, due to the lack of observations and strong observational bias, it is not yet clear whether there is a significant difference in population between the leading and trailing islands \citepads{Chen2019}.

\citetads{Tiscareno2009} found that the long term stability of Twotinos is weaker than Plutinos, implying that a smaller percentage of Twotinos have been preserved since the formation of the Solar System. Additionally, the primitive population of Twotinos may also have changed because the resonance might capture the scattered objects from Kuiper belt in the 4.5\,Gyr's evolution \citepads{Lykawka2007}. Therefore, it is still an open question how the population in the asymmetric islands obtained during the era of planetary migration has changed.

The presence of high-inclination objects in resonances is another interesting issue, because the resonance capture during the planetary migration favours strongly in trapping of low-inclination objects. Some researchers proposed that the scattering events can account for the existence of high-inclination objects \citepads[e.g.][]{Gomes2003, Levison2008}. However, \citetads{Nesvorny2015b} successfully reproduces the inclination distribution of small objects in slow migration model, in which the secular resonances \citepads[e.g.][]{Milani1990} are believed to play an important role in pumping up  the inclination. In fact, several secular mechanisms such as secular resonance, secondary resonance, and  the von-Zeipel-Lidov-Kozai (ZLK) mechanism, may significantly influence the behaviour of objects inside MMRs \citepads[e.g.][]{Morbidelli2002, Gallardo2012, Saillenfest2016}.

Nevertheless, \citetads{Morbidelli1995} argued that in outer resonances such as the 1:2, 2:5, and 1:3 resonances, the presence of low-order secular resonances was not so evident. This is because the precession rates of these outer objects are much slower than those of major planets. 
However, some secular mechanisms, e.g. the secular resonance related to the precession of Neptune, three-body resonances associated with Uranus \citepads{Nesvorny2001}, and the ZLK mechanism \citepads{Nesvorny2001, Lykawka2007, Tiscareno2009, Li2014b}, have been found inside the 1:2 resonance. 
In addition, even in the absence of low-order secular resonances, 1:N resonances can still exhibit chaotic diffusion due to the presence of multiple resonance centres. Small objects moving in the vicinity of separatrix between different resonance islands may have somewhat irregular orbits, and such chaotic diffusion can make the stable region of the 1:2 resonance relatively fragmented \citepads{Melita2000}.

To reveal the complex and intriguing structure of the 1:2 resonance as well as its significant implications for the evolutionary history of the Solar System, we carry out a thorough investigation on the phase space of the 1:2 resonance. And as a comparison, the 1:3 resonance is also analysed. The rest of this paper is organized as follows. In Section 2, we introduce the methods and dynamical model applied in this paper. Then the secular mechanisms inside the 1:2 and 1:3 resonances are figured out and their dynamical effects are analysed in Section 3. By combining the dynamical features of the 1:2 resonance with previous knowledge of resonance capture, we explain in Section 4 the eccentricity distribution of Twotinos. Finally, our conclusions are presented in Section 5.

\section{Model and Methods}

\subsection{Resonance centre} \label{sec:ResCen}
The critical angle of an eccentricity-type 1:N resonance is $\phi=N\lambda-\lambda_8-(N-1)\varpi$, where $\lambda$ and $\varpi$ represent the mean longitude and the perihelion longitude of the object, while the subscript ``8'' refers to the 8th planet (Neptune) from the Sun as usual. For symmetric resonance at low orbital inclination, $\phi$ librates around the exact solution $0^\circ$ or $180^\circ$. These symmetric resonance solutions lose stability at a certain eccentricity when the asymmetric solutions appear. In the circular restricted 3-body (CR3B) model, these solutions correspond to the minima of disturbing function, and thus can be calculated numerically. The disturbing function can be expressed as an expansion series \citepads[see e.g.][]{Lei2021}, and we adopted the  analytical method introduced by \citetads{Lei2021}, numerically calculated the average of the short-period terms, and determined the location of the resonance centre. We note that the value of $\phi$ at the resonance centre depends on the eccentricity, the inclination, as well as the argument of perihelion. Here for simplicity we set the argument of the perihelion $\omega = 0^\circ$, and calculate the asymmetric centre in the 1:2 resonance in a CR3B model consisting of the Sun, Neptune and a massless twotino, in which the mass ratio (the secondary body's mass to the total mass) $\mu=5.146\times10^{-5}$.  

\begin{figure}[!htb]
\centering
\resizebox{\hsize}{!}{\includegraphics{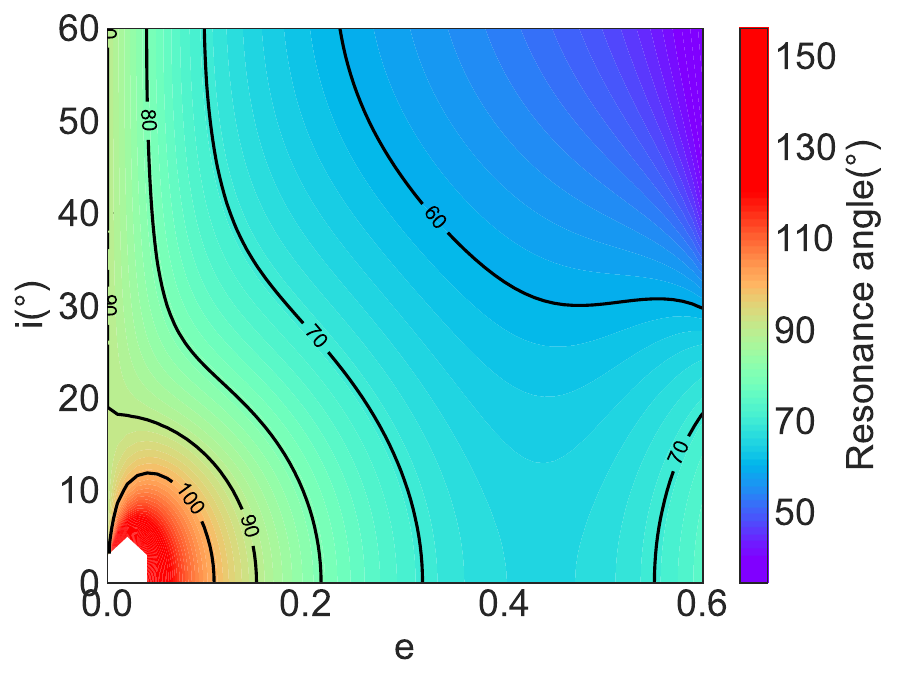}}
\caption{Variation of the resonance centre in the leading asymmetric island of the 1:2 resonance with eccentricity ($e$) and inclination ($i$). The argument of perihelion is set as $\omega=0^\circ$. The colour indicates the resonant angle $\phi$ at the resonance centre. The blank area in the lower left corner indicates that the asymmetric resonance does not exist in that region.}
\label{fig:resonance_center}
\end{figure}

Fig.~\ref{fig:resonance_center} shows how the asymmetric centre varies with eccentricity and inclination in the CR3B model. At low inclination, the $\phi$ at asymmetric centre depends sensitively on eccentricity. On the other hand, its variation with inclination at medium eccentricity (0.15 to 0.4) is quite limited, and most observed Twotinos are found to have eccentricities in this range.

The real resonance motion happens in the Solar System rather than in the CR3B model. Therefore the `resonance centre' is no longer the periodic solution but refers to the motion of which the object experiences the minimal libration amplitude of the resonance angle compared to adjacent orbits in the orbital element space. The dynamical model we adopt in this paper is the outer Solar System that includes the Sun and four major planets, with the initial orbital elements of these planets given in the ecliptic coordinate system at MJD\,59000.5. Due to perturbations from other planets, the asymmetric resonance centre cannot be calculated directly as the minimum of perturbations any longer as in the CR3B model. Instead, we perform some numerical simulations in the outer Solar System model to determine the asymmetric centre statistically. 

As an example, we show our calculations of the resonance centre for objects with eccentricity $e=0.2$ as follows. First, we set test particles in the same plane as Neptune, with their orbital inclination $i$, longitude of ascending node $\Omega$ and mean anomaly $M$ being the same as the ones of Neptune, i.e. $i=i_8$, $\Omega=\Omega_{8}$ and $M=M_8$. The rest two orbital elements, the semimajor axis $a$ and the perihelion argument $\omega$, are then tested in reasonable ranges, $\omega \in (0^\circ, 360^\circ)$ and $a\in (47.0,48.5)$\,au near the resonance. The orbits of test particles were integrated in the outer Solar System for 1\,Myr. The libration amplitudes of resonance angle $\phi$ were monitored during the integration, and the minimum libration will be defined as the resonance centre. It should be noted that the above choice of fixed $M=M_8$ is somewhat arbitrary. An alternative procedure of finding the resonance centre is to vary the fast angle $M$ but fix $\omega$. Our test runs revealed that these two methods produce equivalent results.  

\begin{figure}[!htb]
\centering
\resizebox{\hsize}{!}{\includegraphics{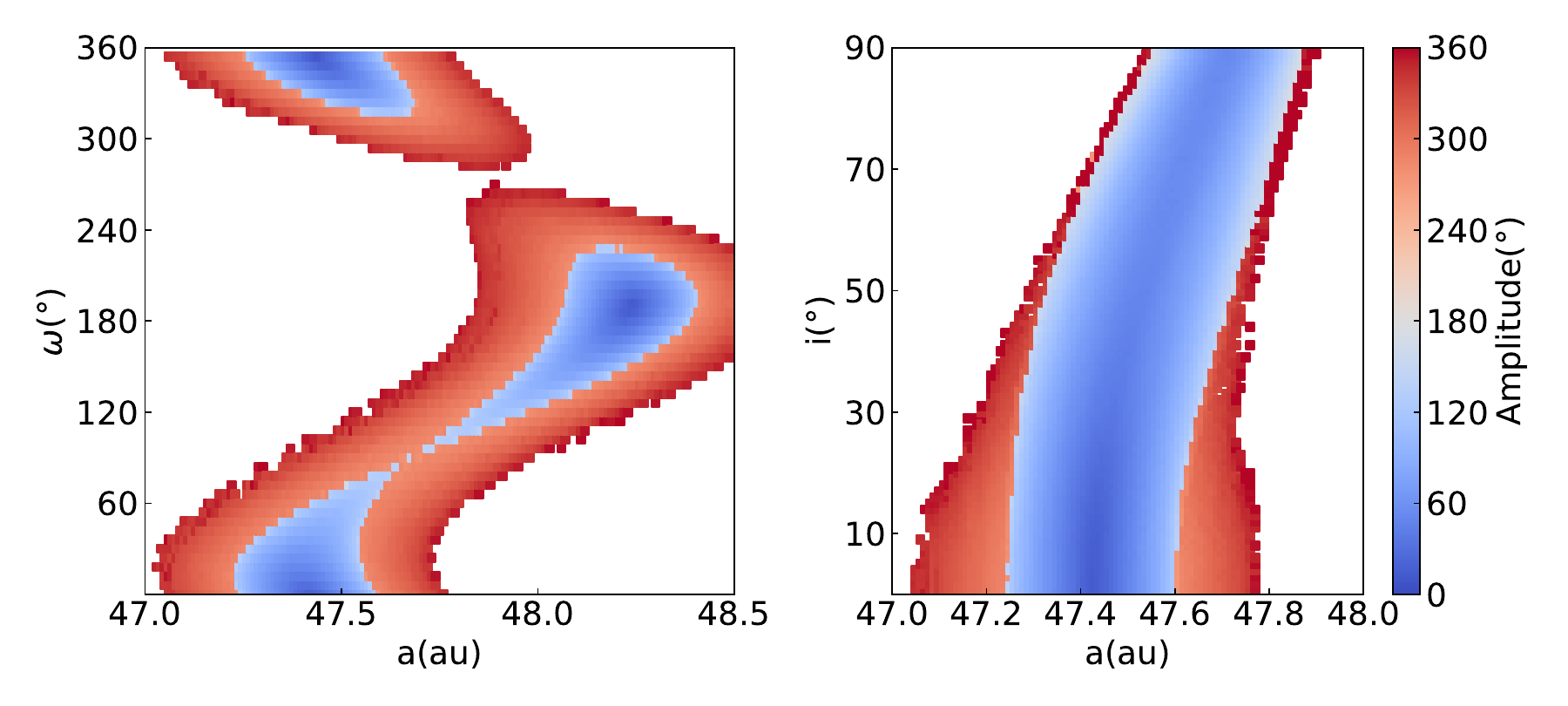}}
\caption{Libration amplitude of resonant angle $\phi$ for initial eccentricity $e=0.2$ on $(a,\omega)$ plane (left) and $(a,i)$ plane (right). The left panel is for coplanar configuration ($i=i_8$, $\Omega=\Omega_{8}$) and the initial mean anomaly is fixed as $M=M_8$. The right panel is for the inclined orbits with fixed initial argument of pericenter $\omega = -3^{\circ}$ (see text). The libration amplitude is indicated by colour. Because the horseshoe resonance configuration encompasses both asymmetric islands, an abrupt change in libration amplitude can be seen at the separatrix.}
\label{fig:prerun}
\end{figure}

We plot the libration amplitudes found in the integrations in the left panel of Fig.~\ref{fig:prerun}. Clearly can be seen, the leading and trailing centres are located at different initial values of $a$ and $\omega$. The leading centre is near $a=47.4$\,au while the trailing one is near 48.2\,au. The displacement in the semi-major axis is primarily caused by the short-period oscillations of planets \citepads[e.g.][]{Nesvorny2001, Zhou2009}. It is worth noting that this displacement occurs when the instantaneous orbital elements are used as the initial conditions, and it does not mean that the leading and trailing islands have different average nominal semi-major axes over an extended period of evolution.  The minimum libration of resonance centres happens correspondingly at  $\omega \approx -3^{\circ}, 177^\circ$, respectively. It must be emphasized that these values of $\omega$ are specific for fixed $M=M_8$, which is an arbitrarily choice. In fact, if we fix $\omega$ and vary $M$, we can find the initial $M$ for any $\omega$ to achieve the minimum libration.

At the leading centre, $\omega \approx -3^{\circ}$ corresponds to an initial resonance angle $\phi=84.4^{\circ}$. We note that this resonance centre is for the coplanar orbits with $e=0.2$. For orbits with other inclination and eccentricity, the resonance centres should be calculated individually following the same method as described above.  

 Based upon the previous results, we conducted further exploration on the $(a,i)$ plane. Adopting $e=0.2, \omega=-3^{\circ}, \Omega=\Omega_{8}, M=M_8$ for test particles and setting their $a$ and $i$ on a $100\times 90$ grid on the $(a,i)$ plane with $a \in (47.0{\rm au}, 48.0{\rm au})$ and $i \in (0^\circ,90^\circ)$, we integrated and monitored their orbits in the outer Solar System for 1\,Myr. The libration amplitudes of these orbits are summarized in the right panel of Fig.~\ref{fig:prerun}.  One may noticed that the resonance structure shows a curvature and the location of the minimal libration amplitude skews outward towards larger $a$ as the inclination increases. The main reason for this deviation of semimajor axis remains the selection of initial orbital elements (mainly the semimajor axis). In fact, for all particles, regardless of their orbital inclinations, the averaged $a$ at the resonance centre is approximately 47.8\,au during the long-term evolution. For the same reason, the trailing island skews inward as the inclination increases. It is worth noting that the quasi 1:2 MMR between Uranus and Neptune may also contribute little (about 0.005\,au in the opposite direction from Fig.~\ref{fig:prerun}) to such deviation \citepads{Zhou2020}. After determining  the initial conditions for resonance centre, our analyses on the motion in the 1:2 MMR with Neptune will be carried out all for objects around the resonance centre.

\subsection{Frequency analysis}
The frequency analysis is often used in studying the long-term orbital evolution of celestial objects \citepads[see the pioneer work in e.g.][]{Laskar1990, Laskar1993, Robutel2001}. The basic idea of the frequency analysis is to obtain the key information about mechanisms that control the long term evolution of objects by integrating their orbits over a short timescale. Through frequency analysis, we can determine the proper frequencies of orbital precessions, which provide insights into the possible secular resonances that an object may experience. The characteristic of the power spectrum calculated from the evolution of certain orbital elements can also tell the regularity of the motion.  The effectiveness of frequency analysis has also been demonstrated in previous studies of main belt asteroids \citepads[e.g.]{Michtchenko1995, Michtchenko2002} and Trojan asteroids of different planets \citepads{Zhou2009,Zhou2011,Zhou2019, Zhou2020,Zhou2021}.

\subsubsection{Simulation parameters}
 
The four major planets in the Solar System have eigenperiods ranging from 46\,kyr to 1.9\,Myr, except for the very slow precession of Jupiter’s ascending node ($\sim$130\,Myr). In the outer Solar System, two quasi MMRs may influence the motion of objects. One is the quasi 5:2 resonance between Jupiter and Saturn (also known as the Great Inequality) and the other is the quasi 2:1 resonance between Uranus and Neptune. The eigenperiods of them are $\sim$880\,yr and $\sim$4200\,yr, respectively. To obtain the information about the long term mechanisms that may affect the motion, the timescale of numerical simulations of the motion should cover these proper periods as much as possible \citepads[see e.g.][]{Nobili1989, Zhou2009}. In this paper, we output the simulation data at an interval of 256\,yr and a total integration time of $2^{25} \approx 3.4 \times 10^{7}$\,yr is adopted. This allows us to distinguish periods ranging from 512\,yr to 17\,Myr and covered most of the eigenperiods in our Solar System, except for the precession of Jupiter’s ascending node.

We used the {\it Swifter\_symba} integrator package \citepads{Levison2000} with an on-line low-pass digital filter module \citepads[see e.g.][]{Michtchenko1993,Michtchenko1995,Zhou2020}. This technique effectively filters out the short-period terms (e.g. planetary mean motion) and minimizes the interference from high frequencies. The ecliptic plane is adopted as the reference plane in our calculation.  We focused on three terms that are tightly related to the long term evolution in our analyses, i.e. $\cos\phi$, $e\cos\varpi$, and $i\cos\Omega$. The proper frequencies of these terms, denoted by $f$, $g$, and $s$ respectively, are just the frequencies of the critical angle of the 1:2 MMR, the perihelion precession, and the ascending node precession. 

For TNOs, the most influential secular resonances are mainly associated with Neptune. Because of their great distance from the Sun, TNOs have relatively low proper frequencies. In the 1:2 resonance, the typical libration timescale of the resonance angles is approximately 10\,kyr to 100\,kyr \citepads[see e.g.][]{Lan2019, Gallardo2020}, while the precessing timescales of their ascending node and perihelion are often on the order of millions of years. For the 1:3 resonance at larger distance, the precession timescales can reach up to 10\,Myr. Therefore, we quadrupled both the output interval and the total integration time when studying the 1:3 MMR. 

Also, for the 1:3 MMR, we ignore the quasi 5:2 resonance between Jupiter and Saturn. Our integrations have demonstrated that it does not yield significant effects because its period is smaller by orders of magnitude than the eigenperiods of TNOs. The quasi 2:1 resonance between Uranus and Neptune however is still included in our analyses.

\subsubsection{Proper frequencies and spectral number}

After integrating the orbits of test particles, we used a fast Fourier transform (FFT) to obtain the frequency spectra, with which we were able to determine the proper frequencies and assess the regularity of the corresponding motion. Generally, the FFT is accurate enough and is very efficient in obtaining power spectra of time series produced by the numerical integrations of orbits. To check the accuracy of our methods, we compare the frequencies of major planets obtained from our calculation with the ones in literature \citepads{Nobili1989,Zhou2020} in Table~\ref{tab:eigenfre}. Our results are in good agreement with that in earlier references. 

\begin{table}
	\caption{Proper frequencies of major planets in references \citepads{Nobili1989,Zhou2020} and in this paper. The frequencies are given in $10^{-4}2\pi\,{\rm yr}^{-1}$. The $g_{5,\cdots,8}$ and $s_{5,\cdots,8}$ stand for the precessions of perihelion and ascending node for planets Jupiter, Saturn, Uranus and Neptune, respectively. The 2N:1U and 5S:2J are for the quasi MMRs between two pairs of planets. }
	\label{tab:eigenfre}
\centering
\begin{tabular}{cccc} 
\hline\hline
Frequency      & Nobili 1989 & Zhou 2020 & This paper  \\ 
\hline
$g_5$    & 0.03285     &  0.03263    & 0.03262             \\ 
$g_6$    & 0.21794     &  0.21353    & 0.21737             \\ 
$g_7$    & 0.02382     &  0.02384    & 0.02385             \\ 
$g_8$    & 0.00519     &  0.00522    & 0.00519             \\ 
$s_5$    & 0.00008     &  0.00008    & \textless 0.00029  \\ 
$s_6$    & 0.20328     &  0.20355    & 0.20305             \\ 
$s_7$    & 0.02309     &  0.02325    & 0.02296             \\ 
$s_8$    & 0.00534     &  0.00536    & 0.00537             \\ 
\hline
2N:1U &  2.3606   &  2.3238    & 2.35524            \\ 
5S:2J  &   N.A.      &  N.A.        & 11.33012          \\
\hline
\end{tabular}
\end{table}

After performing an FFT on the data from numerical simulations of the motion, we can identify the strongest peaks in each frequency spectrum. A `dynamical spectrum' is constructed by plotting the frequencies of the highest peaks in the power spectra for orbits with fixed initial elements but varying $a$ or $i$. Since the proper frequencies generally varies continuously with $a$ or $i$, they can be easily recognised in the dynamical spectrum. This method of picking out the proper frequencies from the dynamical spectrum has been successfully used in our previous work \citepads[see details e.g. in][]{Zhou2009, Zhou2019, Zhou2020}.

It is worth noting that the proper frequencies have not only the magnitude but also the direction. For a non-resonant TNO, the overall perihelion precession is positive, while its precession of ascending node is negative \citepads[e.g.][]{Knezevic1991}. For objects in the 1:2 resonance, although the $\dot{\Omega}$ may sometimes take positive values for highly inclined polar orbits, and the $\dot{\varpi}$ can be positive in extremely chaotic horseshoe resonances, the precession of both $\Omega$ and $\varpi$ is negative for the majority of Twotinos. The same is true for the ascending nodes of major planets (with respect to the invariable plane). However, for the planets' perihelion, the precessions are all positive. Since the perihelion precession rate $g$ is in the opposite direction to $g_8$, we would not find the $\nu_8$ resonance where ($\varpi-\varpi_8$) librates. For similar reasons, any other secular resonances involving $g_5$, $g_6$, $g_7$, or $g_8$ are also unlikely to exist.

Besides the frequencies, the spectrum may also tell us the stability of the orbits. In fact, we can calculate from a power spectrum the spectral number (SN) and  use it as an indicator of orbital stability \citepads[for more details, see e.g.][]{Michtchenko1995, Zhou2009, Zhou2019}. The SN is defined as the number of peaks with amplitudes above a certain threshold in a frequency spectrum. Specifically, in this paper we adopt a threshold of 1\% of the highest peak in each spectrum. A small SN indicates a `clean' spectrum and thus a regular motion, while a large SN implies a noisy spectrum and thus irregular orbits and the onset of chaos. 

\section{Dynamics of 1:2 and 1:3 MMRs}
Knowing the location of the resonance centre at given eccentricity and inclination, we can explore the dynamics of the resonance around the resonance centre. For example, for a given eccentricity $e$, we can set the initial conditions of test particles on a grid in the $(a,i)$ plane, and the rest orbital elements except for $(a,e,i)$ are assigned the values at the corresponding resonance centre. The orbits of these test particles are integrated in the outer Solar System and then analysed. We note that as for the dynamical properties, the leading resonance island is absolutely identical to the trailing island, therefore it is enough to analyse only one of them (the leading one in this paper) arbitrarily.

\subsection{1:2 MMR in $(a,i)$ plane}

\subsubsection{Maps of proper frequencies}

\begin{figure}[!htb]
\centering
\resizebox{\hsize}{!}{\includegraphics{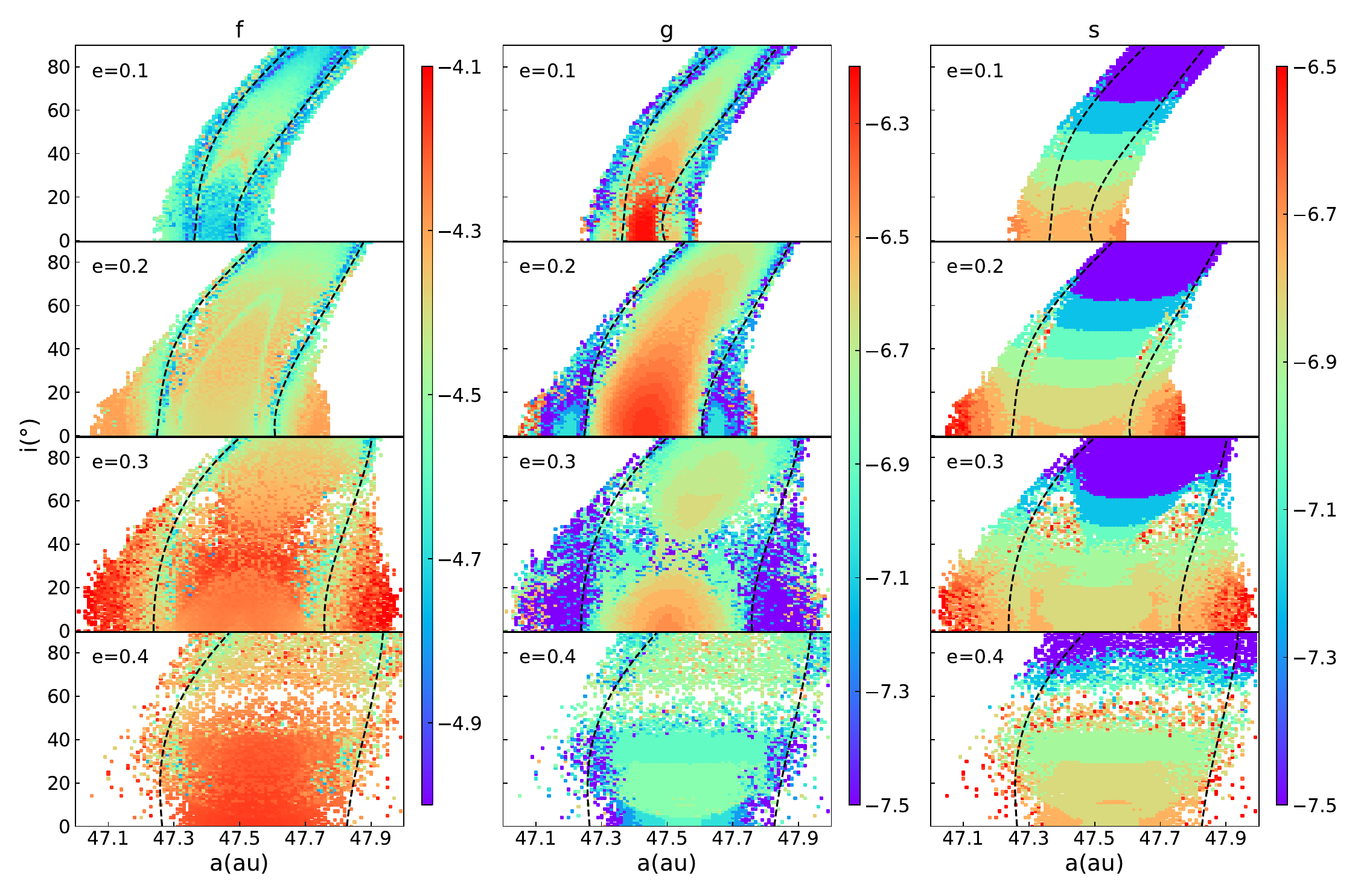}}
\caption{Proper frequencies of test particles' motion in the 1:2 MMR on $(a,i)$ plane. From left to right, the panels show the proper frequencies of the resonance angle ($f$), of perihelion ($g$), and of ascending node ($s$), respectively. From top to bottom, test particles have increasing initial eccentricities, from 0.1 to 0.4, as labelled in each panel. The colour indicates the logarithm of the proper frequency in $2\pi\,{\rm yr}^{-1}$. The black lines mark the boundary between the horseshoe and asymmetric resonance islands. In between the lines are the asymmetric resonance island. In the blank area, orbits initialised there are unstable and cannot survive the orbital integration of $\sim$34\,Myr. }
\label{fig:eigenfre}
\end{figure}

For initial eccentricities $e=0.1, 0.2, 0.3$ and $0.4$, we set thousands of initial conditions on $(a,i)$ plane and integrated the orbits. The proper frequencies, $f$ of the libration of resonance angle ($\phi$), $g$ of the precession of perihelion longitude ($\varpi$) and $s$ of the precession of the ascending node ($\Omega$), are calculated and shown in Fig.~\ref{fig:eigenfre}. The boundary between horseshoe resonance and asymmetric resonance is delineated in the numerical simulations, and the black lines in the figure represent a polynomial fitting of such boundary. We note that in Fig.~\ref{fig:eigenfre} and all subsequent dynamical maps, the semi-major axis, eccentricity, and orbital inclination are the initial values (osculating elements), and these orbital elements will change during subsequent evolution.

As the perturbation theory \citepads[see e.g.][]{ssd} predicts, for small $(e,i)$ orbits, the proper frequency $g$ ($s$) decreases with increasing eccentricity (inclination). In Fig.~\ref{fig:eigenfre} we also see that the proper frequency $f$ increases with increasing eccentricity, which reflects a stronger resonance strength in the resonance centre at higher eccentricity. On the other hand, the fastest precession of perihelion tends to occur in the centre of the asymmetric island, as long as the eccentricity does not exceed 0.3. The frequency $f$ is relatively lower near the boundary between horseshoe and asymmetric resonance islands. In fact, the period at the (ideal) separatrix is infinite, but in the outer Solar System model, various perturbations from other planets blur the boundary and particles nearby might switch between different resonance modes in their evolution.

There are some other interesting structures in Fig.~\ref{fig:eigenfre}, such as the arch curve in $f$ when $e=0.1$ and $e=0.2$, and the gap structure at high inclination in all $f, g, s$ when $e=0.3$ and $e=0.4$. These structures are related to secular mechanisms that will be discussed below.

\subsubsection{Stability maps and secular mechanisms}

\begin{figure}[!htb]
\centering
\resizebox{\hsize}{!}{\includegraphics{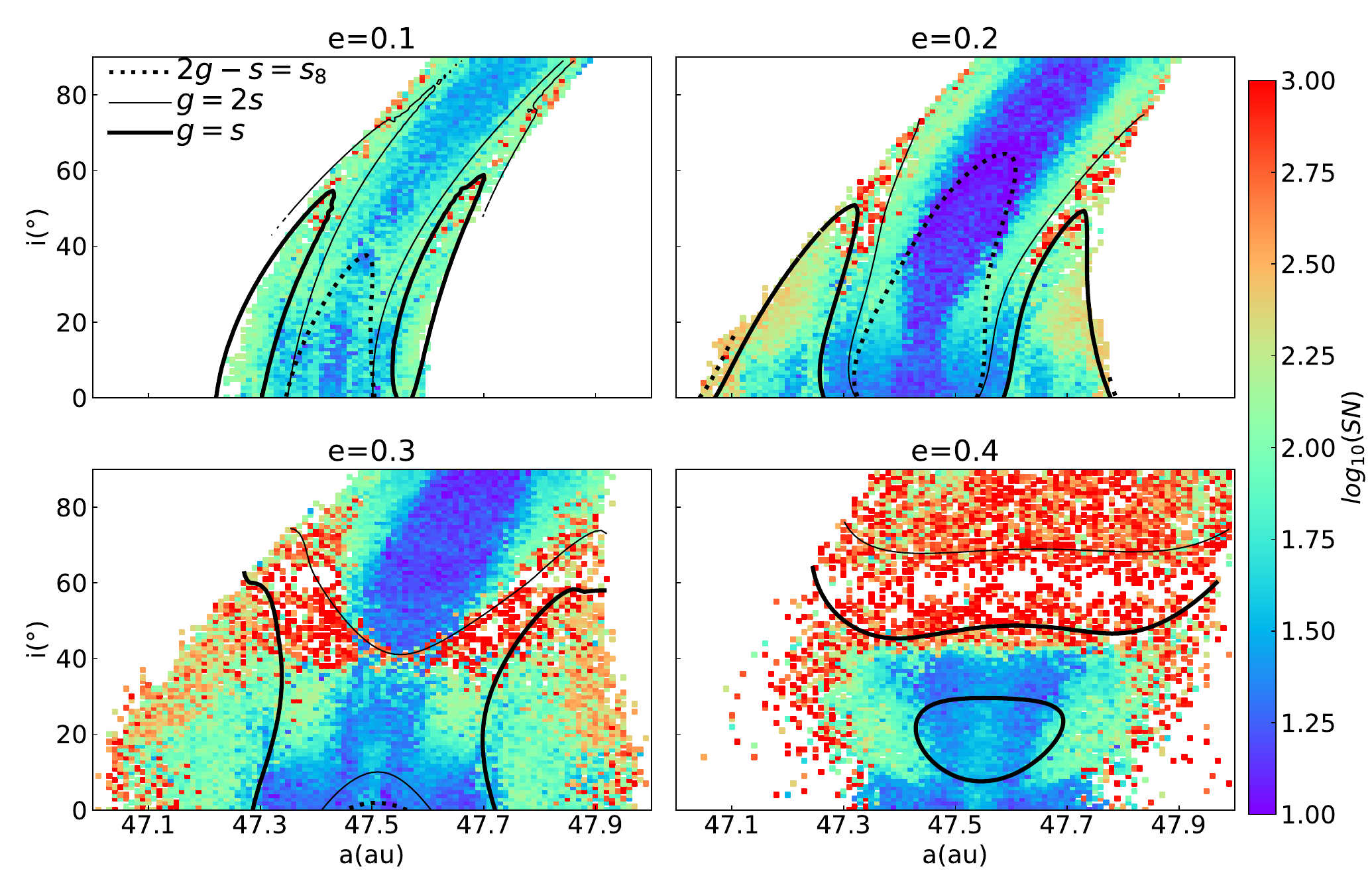}}
\caption{Maps of stability of orbits in the 1:2 MMR. The colour represents the logarithm of the SN calculated from $\cos\phi$, with blue indicating regularity and thus stability while red for instability. Lines of different styles are used to indicate the positions of various secular mechanisms (see text).}
\label{fig:sn}
\end{figure}

We use the SN to indicate the regularity of orbits. Although an irregular orbit is not necessarily unstable, the SN still reveals the overall stability of orbits. The maps of SN (calculated from $\cos\phi$) for four initial eccentricities are illustrated in Fig.~\ref{fig:sn}. According to the stability maps, the most stable orbits in the 1:2 MMR locate around the asymmetric resonance centre. For orbits with small eccentricities ($0.1, 0.2$) the most stable region in the ($a,i$) plane extends from $i=0^\circ$ (coplanar with Neptune) to $i=90^\circ$. A gap of instability at $i\sim40^\circ$ appears when $e=0.3$ and it expands to high-inclination region when $e=0.4$.  In addition, the resonance width in semimajor axis is much wider when $e=0.2, 0.3$ than when $e=0.1$, and generally the width gets smaller as the inclination increases. Some other fine structures in the maps, for example, vertical stripes of less regular orbits (relatively larger SN) in the low inclination region at $e=0.2$ and very regular orbits of low inclination ($i\lesssim 10^\circ$) when $e=0.2, 0.3, 0.4$, can be seen in Fig.~\ref{fig:sn}. The mechanisms that produce these structures will be figured out later. 

The proper frequencies $f, g, s$ have been calculated (Fig.~\ref{fig:eigenfre}), and they can be regarded as functions of  the orbital parameters ($a, e, i$). We fitted the calculated $f, g, s$ as functions of $a,e,i$ by polynomials, and used these polynomial functions to determine the locations of specific resonances (simply defined as the equality between frequencies and/or their combinations) on the representative plane, e.g. $(a,i)$ plane.  Such technique of detecting resonances has been applied in our previous work \citepads[please refer to e.g.][for details]{Zhou2009, Zhou2019}. The contour curves on Fig.~\ref{fig:sn} obtained in this way indicate the locations where three prominent secular mechanisms occur. As labelled in Fig.~\ref{fig:sn}, the contours indicate the locations where $g=s$ (ZLK mechanism), $g=2s$, and $2g-s=s_{8}$.

The primary cause of instability for test particles is the ZLK mechanism $g=s$ \citepads[for a review see]{Kozai-rev}, which results in large oscillations in eccentricity through the exchange of eccentricity and inclination. The high eccentricity brings the perihelion of test particles too close to the region where planets may strongly influence, and this is the most immediate reason for test particles to fall out of MMR. Specifically, for objects in the 1:2 MMR, the critical eccentricity is approximately 0.37 for a Neptune-crossing orbit and about 0.6 for a Uranus-crossing orbit.

It should be noted that the actual region affected by ZLK mechanism extends far beyond the solid line in Fig.~\ref{fig:sn}. Empirically, if the difference between $g$ and $s$ is less than approximately $10^{-7}$\,$2\pi\,{\rm yr}^{-1}$, the ZLK mechanism is very likely to occur. The actual frequencies of a test particle are not constant but fluctuate slightly around the nominal values. This phenomenon is referred to as the `frequency drift', which has previously been observed in the inner Solar System and been believed to be accountable for the chaotic nature of the inner Solar System \citepads{Laskar1989,Laskar1990}.

At low eccentricity, the  ZLK region roughly coincides with the boundary of horseshoe resonance island. When $e=0.1$, almost all particles in horseshoe resonance are subject to the ZLK mechanism. At $e=0.2$, the expansion of the horseshoe island allows some particles deep inside it to remain unaffected by the ZLK mechanism, corresponding to the blue area at low inclination. As previously mentioned, particles near the boundary frequently switch between resonance modes under the perturbation of major planets. Superimposed with the separatrix between horseshoe and asymmetric resonance modes, the ZLK mechanism introduces even more irregularity to the motion and the corresponding SN gets larger. However, due to their low initial eccentricity, most particles here do not reach eccentricities high enough to destabilize their orbits even under the influence of the ZLK mechanism.

The ZLK mechanism influences much larger area in the $(a,i)$ plane as the eccentricity increases. When $e=0.3$, in addition to the boundary region between different resonance islands, the ZLK mechanism also occurs within the asymmetric islands and forms an unstable gap at $i\sim40^\circ$, while a stable area remains at higher inclinations. When the eccentricity reaches 0.4, the  ZLK region dominates almost the entire resonance region from $i=10^\circ$ to $70^\circ$. The most notable pattern is an unstable gap near $i=60^\circ$, which is so chaotic that few particles survive the numerical simulation. \citetads{Gallardo2012} suggest that for bodies within resonance, substantial variations in the perihelion distance due to the ZLK mechanism only occur when the inclination exceeds a certain minimum value (approximately $15\circ$), which aligns with the phenomenon depicted in Fig.~\ref{fig:sn}.

Another interesting feature in Fig.~\ref{fig:sn} is associated with the mechanism of $g=2s$ indicated by the dashed line. Similar to the ZLK mechanism,  $g=2s$ is not in compliance with the D'Alembert's rule and cannot be classified as a secular resonance. In Fig.~\ref{fig:sn}, the most unstable motion (red colour) always occurs in the region in between the lines of $g=s$ and  $g=2s$ (where $s<g<2s$), although a certain inclination ($\sim$40$^\circ$) is needed to trigger the instability. In this region, $g$ is slightly larger than $s$, making it easy for particles to fall into the ZLK mechanism under the influence of frequency drift. The $g=2s$ can also be written as $\dot{\omega}+\dot{\Omega}=2\dot{\Omega}$, so that $g=2s$ means that the ascending node and argument of perihelion precess at the same rate ($\dot{\omega}=\dot{\Omega}$). The mechanism of $g=2s$ acts like a weakened version of the ZLK mechanism, triggering relatively small exchanges of eccentricity and inclination. Another potential effect of the $g=2s$ is to provide protection for particles with $g>2s$ from falling into the ZLK mechanism. As a result, when $e=0.3$ (Fig.~\ref{fig:sn}), regions enclosed by the $g=2s$ line are more stable, even at very high inclinations.

\begin{figure*}[!htb]
	\centering
		\subfigure{
			\begin{minipage}[b]{0.45\textwidth}
				\centering
				\includegraphics[width=\textwidth]{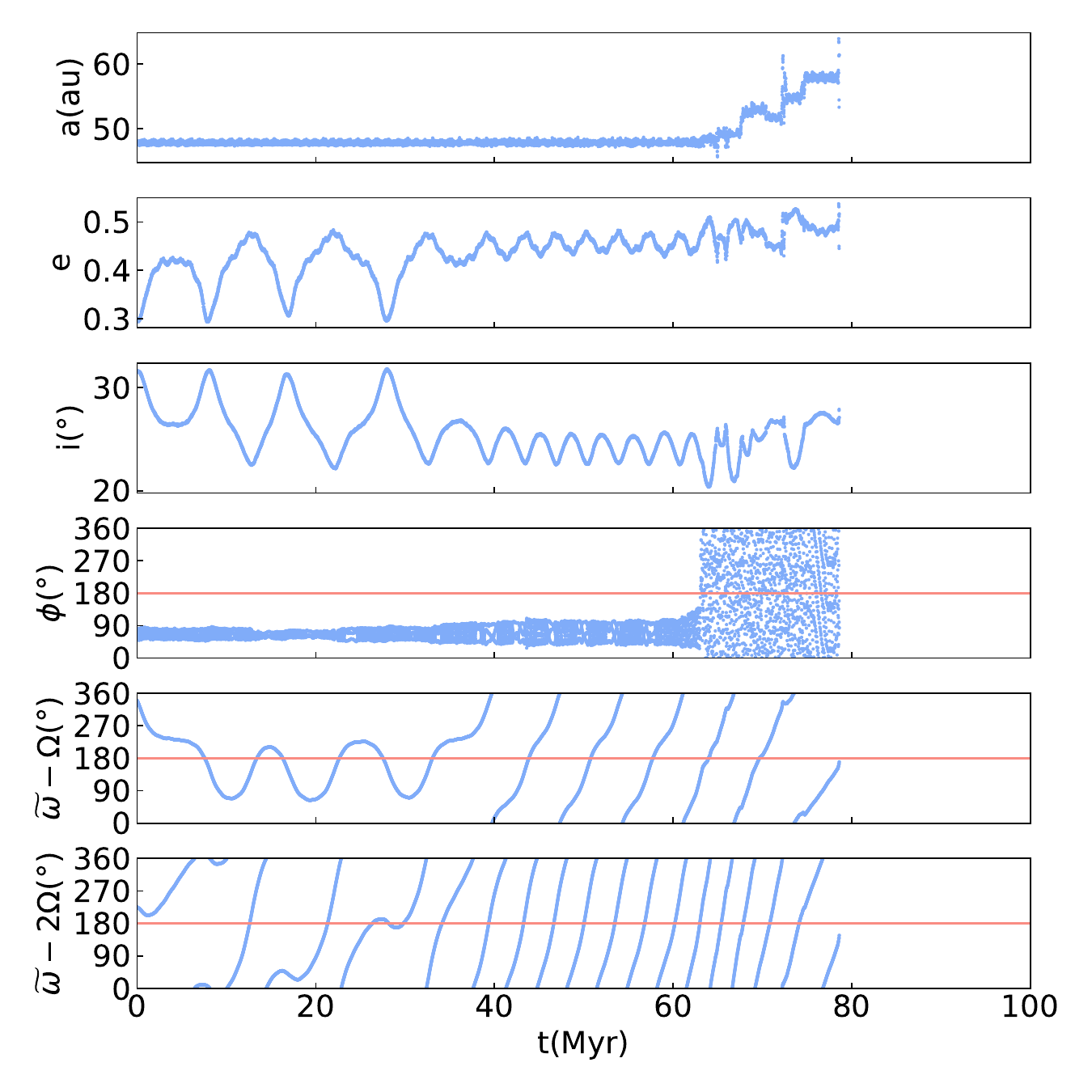}
				\label{fig:Kozaiexample}
			\end{minipage}
		}
			\subfigure{
				\begin{minipage}[b]{0.45\textwidth}
					\centering
					\includegraphics[width=\textwidth]{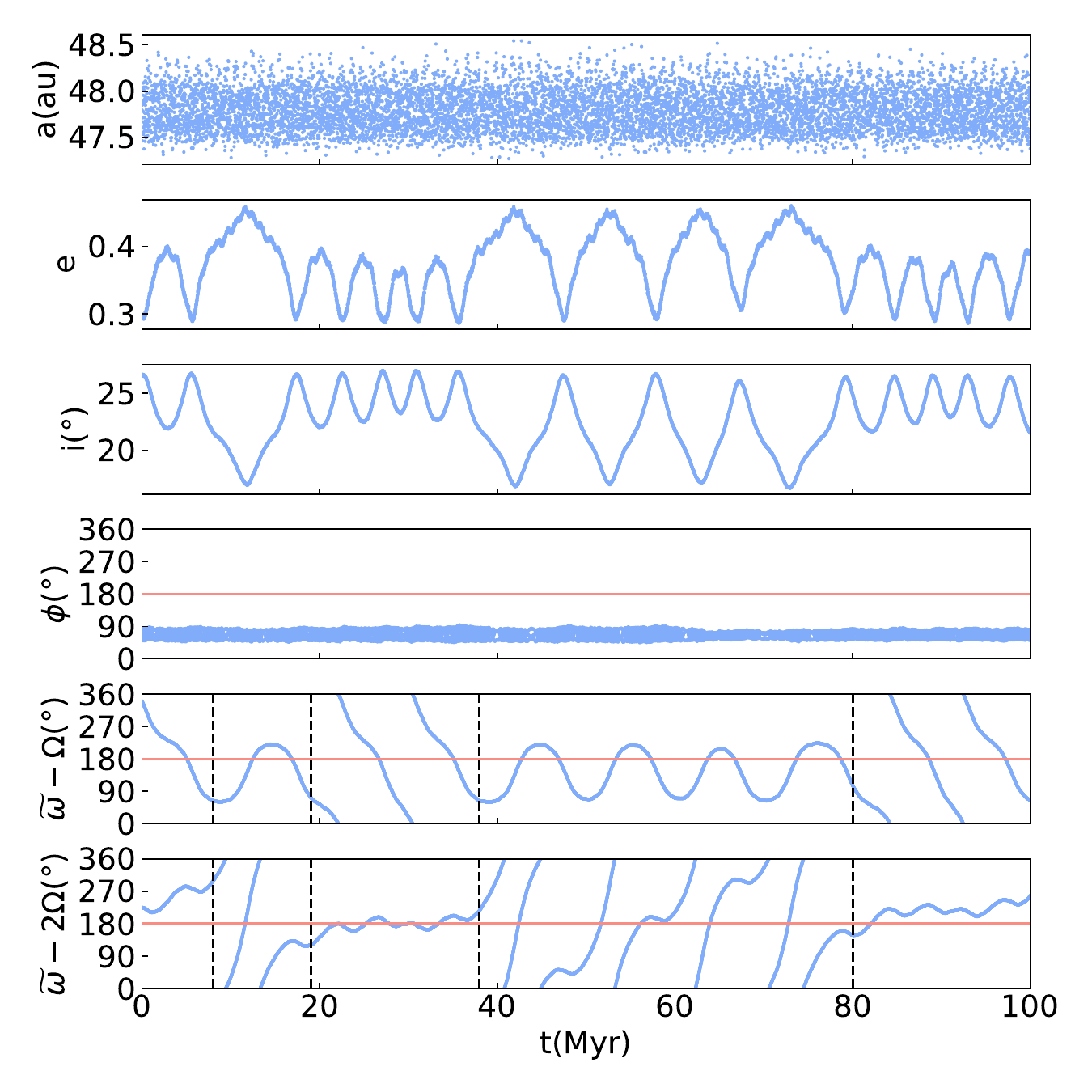}
					\label{fig:g=2sexample}
				\end{minipage}
			}
\caption{Evolution of two typical test particles experiencing the ZLK mechanism (left), and both  ZLK and $g=2s$ mechanisms (right). Two orbits have different initial inclinations $i=33^{\circ}$ (left) and $i=28^{\circ}$ (right),  and the rest initial orbital elements for both orbits are the same ($a=47.5$\,au, $e=0.3$, $\Omega=\Omega_8$, $\omega=342^{\circ}$, and $M=M_8$). From top to bottom, the panels show the evolution of semimajor axis ($a$), eccentricity ($e$), inclination ($i$), the 1:2 resonance angle ($\phi$), the angle of ZLK mechanism ($\varpi-\Omega$), and the angle of $g=2s$ mechanism ($\varpi-2\Omega$). The solid lines indicate the position of $180^\circ$, and the dashed lines mark the times when the orbit switches between two mechanisms.}
			\label{fig:example}
\end{figure*}

To show the dynamical effects of the ZLK mechanism and the mechanism of $g=2s$, we illustrate two typical orbits in Fig.~\ref{fig:example}. As shown in Fig.~\ref{fig:Kozaiexample}, when the ZLK mechanism occurs from $\sim$5 to $\sim$35\,Myr as indicated by the libration of angle $(\varpi - \Omega)$, the test particle’s eccentricity oscillates largely between 0.3 and 0.5 coupling inversely with inclination between $\sim$30$^\circ$ and $\sim$23$^\circ$. The orbit leaves the ZLK mechanism at $\sim$35\,Myr with its eccentricity remaining approximately at 0.5  for tens of millions of years. The high eccentricity, combined with circulation of the perihelion argument, greatly increases the risk of close encounter with Neptune. The test particle escapes the 1:2 MMR at $\sim$63\,Myr and is completely scattered out of the Solar System at around 78\,Myr.

Fig.~\ref{fig:g=2sexample} shows another typical orbit that experiences both the ZLK mechanism and the $g=2s$ mechanism. Over 100\,Myr, the particle switches between the  ZLK and $g=2s$ mechanisms several times. When the critical angle of one mechanism is in libration, the critical angle of the other mechanism is in circulation. The exchange between inclination and eccentricity oscillates with large amplitude during ZLK mechanism phase and it oscillates with moderate amplitude during the phase of $g=2s$ mechanism. 
When $g=2s$ occurs, its critical angle librates around $180^\circ$ with small amplitude. We note that this does not imply that $180^\circ$ is the `centre' of this critical angle because it is in fact related to the selection of coordinate frame. Thanks to the protection provided by the $g=2s$, the particle in Fig.~\ref{fig:g=2sexample} spends less time in a high eccentricity state over 100\,Myr and has a longer lifetime than the particle in Fig.~\ref{fig:Kozaiexample}.

\begin{figure}[!htb]
	\centering
	\resizebox{\hsize}{!}{\includegraphics{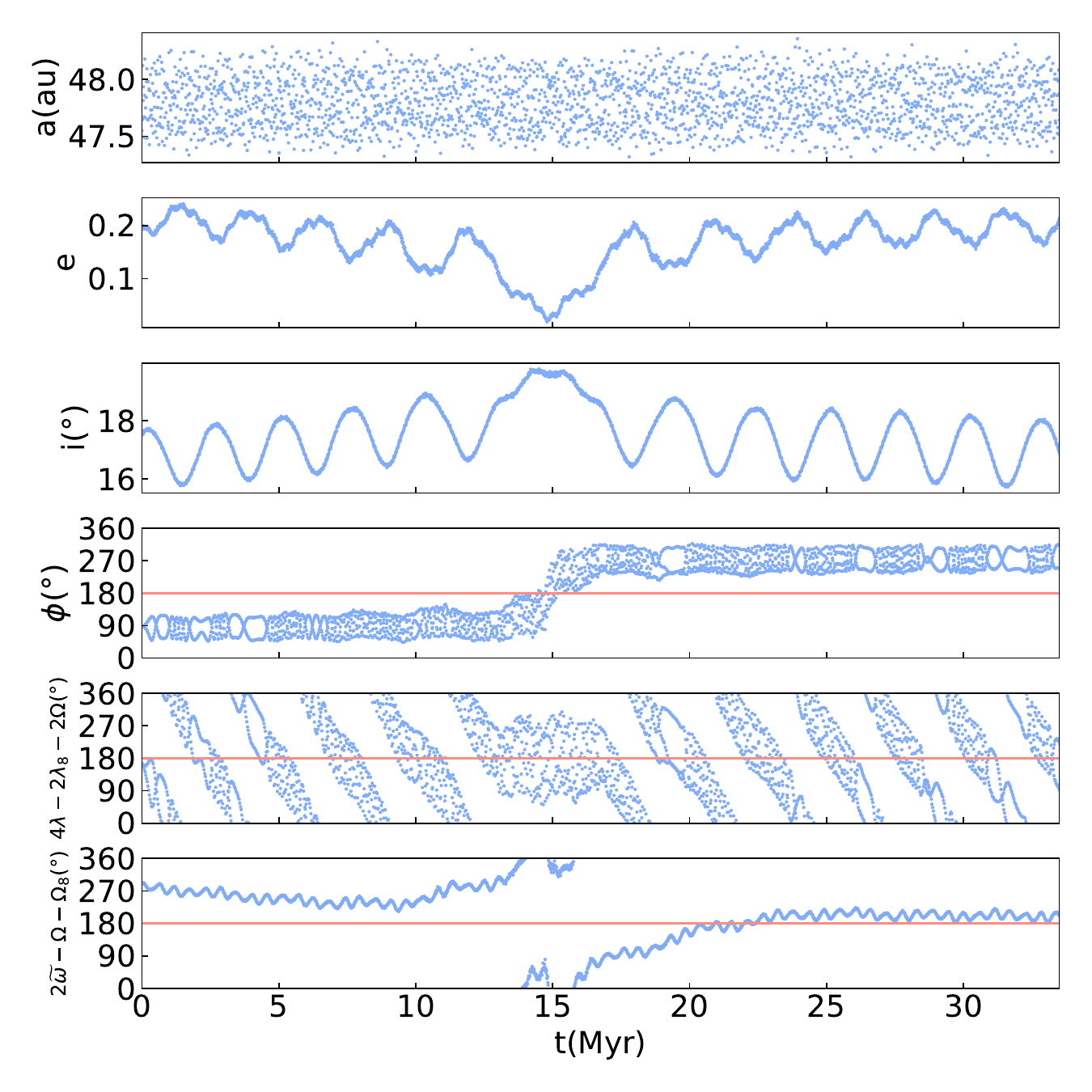}}
	\caption{Evolution of a test particle that experiences the secular resonance $2g-s=s_8$. The initial orbital elements are $a=47.55$\,au, $e=0.2$, $i=19^{\circ}$, $\Omega=\Omega_8$, $\omega=358.86^{\circ}$, and $M=M_8$. From top to bottom, the panels show the semi-major axis, eccentricity, inclination, critical angles of the 1:2 eccentricity-type resonance, of the 2:4 inclination-type resonance, and of the $2g-s=s_8$ secular resonance. The solid lines indicate the position of $180^\circ$.}
	\label{fig:example2}
\end{figure}

In addition to the aforementioned two mechanisms, the secular resonance $2g-s=s_8$ with the critical angle $(2\varpi-\Omega-\Omega_{8})$ occurs within the asymmetric island of the 1:2 MMR. Its location (Fig.~\ref{fig:sn}) coincides with the arch structure in the stability maps of $e=0.1, 0.2$, but it is hardly visible at  $e=0.3$ and it does not appear at $e=0.4$. To show the effect of this secular resonance, we show in Fig.~\ref{fig:example2} an orbit affected by it. Superimposed over the relatively short term ($\sim$2.5\,Myr) variation of inclination (and eccentricity) that is obviously correlated with the variation of critical angle $(4\lambda-2\lambda_8-2\Omega)$, we can find in Fig.~\ref{fig:example2} a relatively long term (a little longer than the integration time of 34\,Myr) variation of inclination and eccentricity, which is correlated with the critical angle ($2\varpi-\Omega-\Omega_{8}$) of this secular resonance. Specifically in this example, the libration of this critical angle is interrupted by the switching of the mean motion resonances (eccentricity-type to inclination-type, and leading island to trailing island), but such libration of ($2\varpi-\Omega-\Omega_{8}$) and its correlation with the behaviour of ($e,i$) can be easily found in neighbouring orbits, from which we know the complete libration period is about 40\,Myr. This is consistent with the results in \citetads[][Figure 9]{Nesvorny2001}.  Similar to the ZLK mechanism, $2g-s=s_8$ increases the amplitude of exchange between eccentricity and inclination. When the particle’s eccentricity is reduced to as low as 0.02, the asymmetric islands of the 1:2 eccentricity-type resonance fade out, and the system recovers the symmetric configuration. Meanwhile, as the inclination increases ($\sim$20$^\circ$), an inclination-type 2:4 resonance takes place. This process lasts for millions of years centring around $t=15$\,Myr in Fig.~\ref{fig:example2}. It is not until the eccentricity value is restored that the 1:2 eccentricity-type resonance resumes the asymmetric islands, by which time the particle has moved from the leading island to the trailing island. 


\subsubsection{Minimal resonance angle and lifespan}

So far, we adopt the SN as the indicator of orbital stability, which is found to be tightly related to the libration amplitude of the resonance angle. Additionally, the libration amplitude can directly reflect which resonance configuration (symmetric, asymmetric, or horseshoe libration) the test particle is in. However, considering the abrupt jump in amplitude when a particle switches between different libration modes, instead of the amplitude we use the minimum of the resonance angle $\phi_{\rm min}$ as a new indicator of orbital stability. The $\phi_{\rm min}$ is defined as the minimal value that can be reached by $\phi$ in the integration of 34\,Myr. We note that this definition of $\phi_{\rm min}$ works well for both the asymmetric librator on tadpole-like orbits around the leading resonance island and the librator on horseshoe-like orbits. But for the motion around the trailing island, which is not considered in this paper, the minimum of $(360^\circ - \phi)$ is the equivalence to $\phi_{\rm min}$. Theoretically, when the eccentricity is very small, there are symmetric orbits that exhibit small-amplitude libration around $180^\circ$, while at high eccentricities, symmetric orbits librating around $0^\circ$ with small amplitude exist \citepads[see e.g.][]{Lan2019}. But these orbits occur so rarely in our simulations that we ignore them here. Under this definition of $\phi_{\rm min}$, a librator with large amplitude will have a small $\phi_{\rm min}$, and on the contrary a particle close to the resonance centre with small libration amplitude will possess a relatively large $\phi_{\rm min}$. 

\begin{figure}[!htb]
\centering
\resizebox{\hsize}{!}{\includegraphics{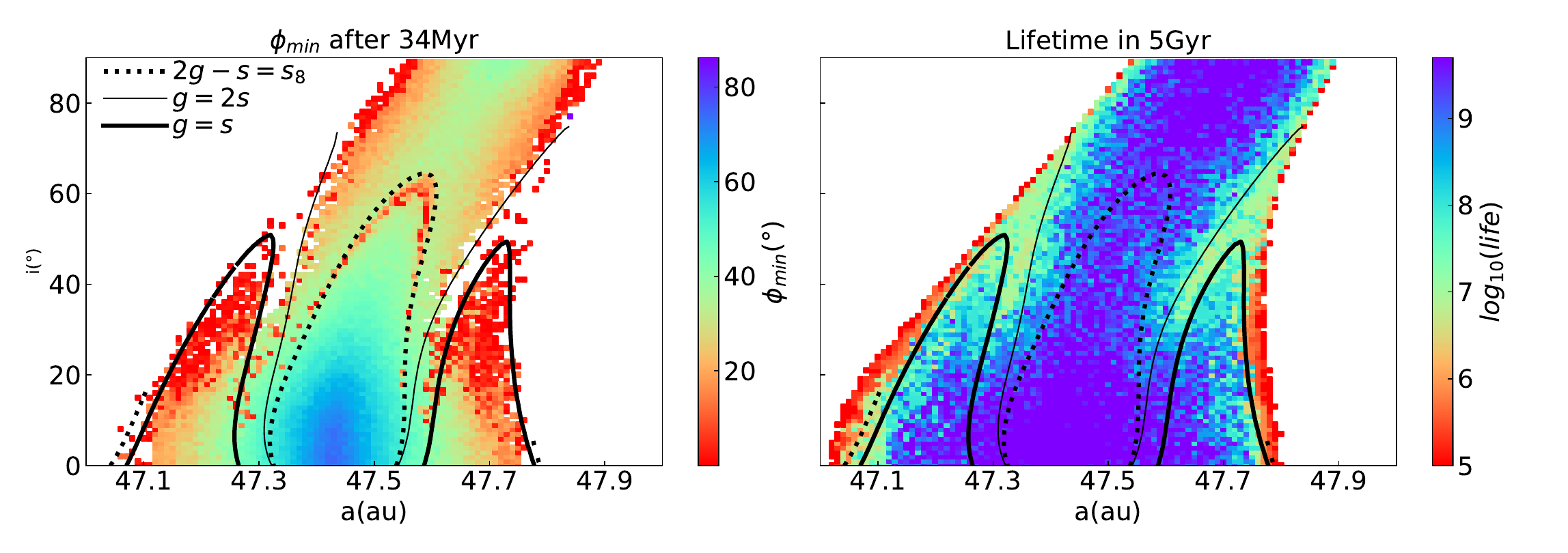}}
\caption{The minimum of resonant angle ($\phi_{\rm min}$) and the lifespan of test particles (see text). Both $\phi_{min}$ (in degrees) and lifespan (in logarithm of years) are indicated by colour. The locations of secular mechanisms are plotted as in Fig.~\ref{fig:sn}.}
\label{fig:lifetime}
\end{figure}

The lifespan of a test particle staying inside the 1:2 MMR is a straight measure of the orbital stability. To further explore the dynamics of the resonance, as well as to check the reliability of the stability indicators, we conducted a long term simulations for particles of $e=0.2$ initialized on the $(a,i)$ plane. The orbits of these particles are integrated in the outer Solar System model up to the Solar System's age (5\,Gyr).  
Together with the $\phi_{\rm min}$ of these orbits calculated in the short-term integration of 34\,Myr, we summarise the lifespans obtained from the 5\,Gyr's integration in Fig.~\ref{fig:lifetime}. 
 
The minimum of resonance angle ($\phi_{\rm min}$) in Fig.~\ref{fig:lifetime} reveals the resonance amplitude very well. Meanwhile, indicating by $\phi_{\rm min}$, the change from asymmetric resonance to horseshoe resonance is smooth, and the abrupt change in amplitude (as shown in Fig.~\ref{fig:prerun}) is avoided. Apparently, $\phi_{\rm min}$ deep inside the asymmetric island is larger, implying relatively stable motions therein. And interestingly, the $\phi_{\rm min}$ around the location of secular resonance $2g-s=s_8$ decreases noticeably because this secular resonance oscillates the eccentricity and triggers switches between asymmetric islands as we have shown in Fig.~\ref{fig:example2}. 

The lifespan map in Fig.~\ref{fig:lifetime} also shows the structure associated with secular mechanisms. The bluest points represent particles that survive the orbital integration up to the age of the Solar System. Generally, after 5\,Gyr's evolution, the area in which particles still retain is significantly reduced. The loss of particles along the edge of stability region can be attributed to chaotic diffusion, while the loss of particles along the  ZLK and $g=2s$ mechanisms is due to the frequency drift as we mentioned before. Within the asymmetric islands, particles escape from the resonance on a gigayear timescale, mainly from the region with inclination $30^\circ$--$70^\circ$. This might be a result of the combined dynamical effects of the secular resonances like the $2g-s=s_8$ and the frequency drift.

\begin{figure}[!htb]
\centering
\resizebox{\hsize}{!}{\includegraphics{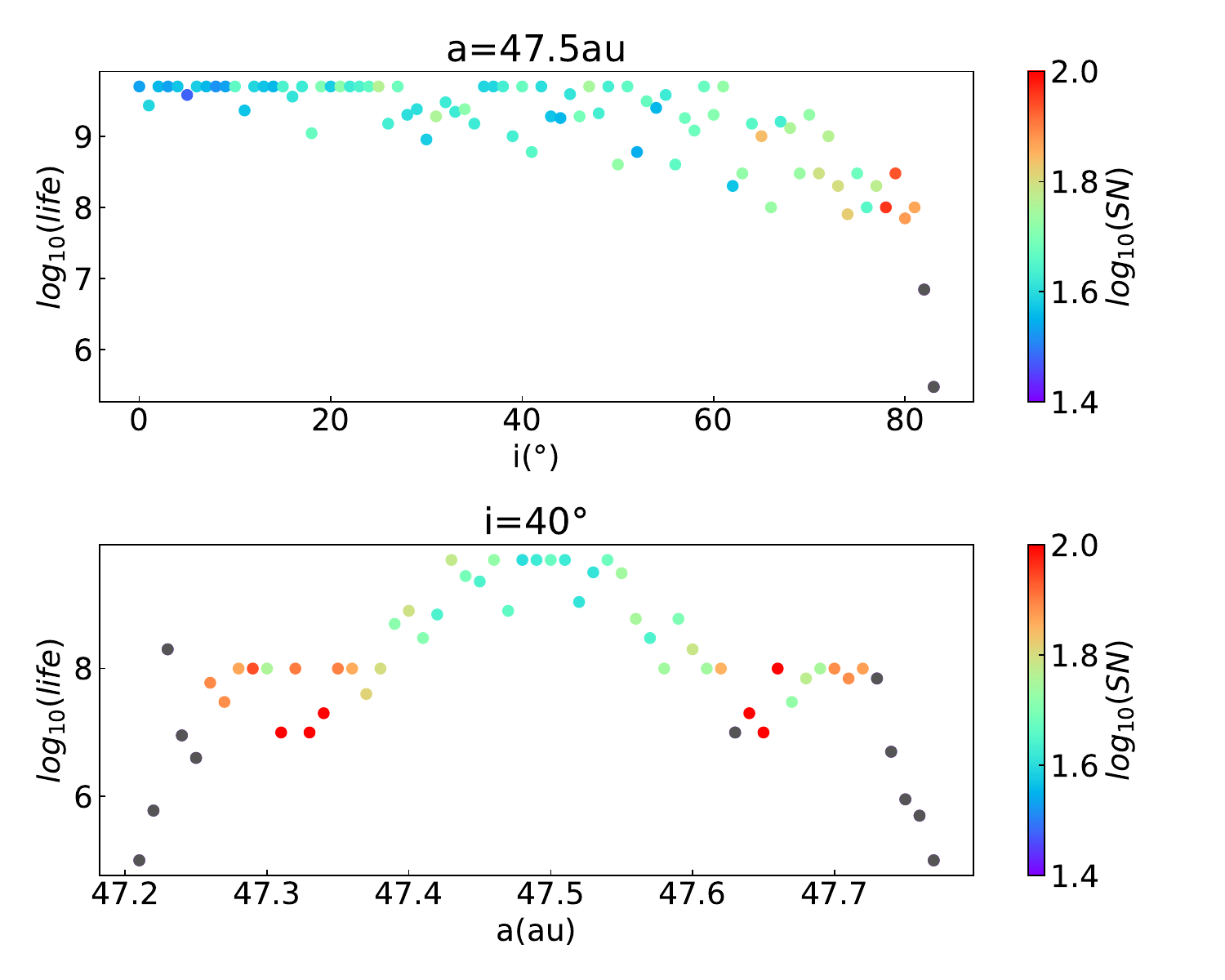}}
\caption{The lifetime and the SN of test particles. Initial conditions are selected from two cross lines in the $(a,i)$ plane, one with fixed $a$ (top) and the other with fixed $i$ (bottom). The colour indicates the value of SN, with those gray dots representing those orbits that are destabilized within the 34\,Myr's integration and therefore the SN cannot be calculated (but the lifetime can still be calculated). }
\label{fig:lifesn}
\end{figure}

We have used the SN as an indicator of orbital regularity, and the regularity to some extent is believed to be equivalent to stability. Since we have obtained the lifespan of particles in the $(a,i)$ plane of $e=0.2$, we can verify such equivalence by comparing the lifespan with corresponding SN value. We select randomly two cross sections in the $(a,i)$ plane, one with fixed semi-major axis $a=47.5$\,au and the other one with fixed inclination $i=40^\circ$, and plot the SN and lifetime in Fig.~\ref{fig:lifesn}. Overall, particles with higher SN have shorter lifetimes. This demonstrates that the SN can be used to estimate the lifetimes of small objects \citepads[see previous similar calculation][]{Zhou2009}. 

To estimate the dynamical lifetime of small objects is very important for understanding the evolution of the Solar System, but it requires massive computational resources to do so through direct integration. It is always of particular interest to find suitable indicators that can be obtained by orbital integration as short as possible. As we have shown above, both the $\phi_{\rm min}$ and the SN can serve as stability indicators, and they overall agree with each other. We will use these indicators below in this paper. 

\subsection{1:2 MMR in $(a,e)$ plane}

We have shown the stability map on the $(a,i)$ plane with 4 specific eccentricities. To check the dependence of the orbital stability on the eccentricity, we construct three dynamical maps on the $(a,e)$ plane, with initial inclinations of $i=i_8$ (coplanar case), $i=20^{\circ}$ and $i=40^{\circ}$, respectively. The values of $a$ and $e$ are chosen to cover the resonance region in steps of 0.01\,au (for $a$) and 0.01 (for $e$), respectively. The rest orbital elements are set as before, $\Omega=\Omega_8$,  $M=M_8$ and $\omega$ are determined in test runs using a similar approach as described in Section~\ref{sec:ResCen}, to place the test particles near the resonance centre. It is worth noting that we also tried to assign specific initial values of $\omega$ according to each initial eccentricity and found this change of initial condition brought no qualitative difference in the results.

\begin{figure*}[!htb]
\centering
\resizebox{\hsize}{!}{\includegraphics{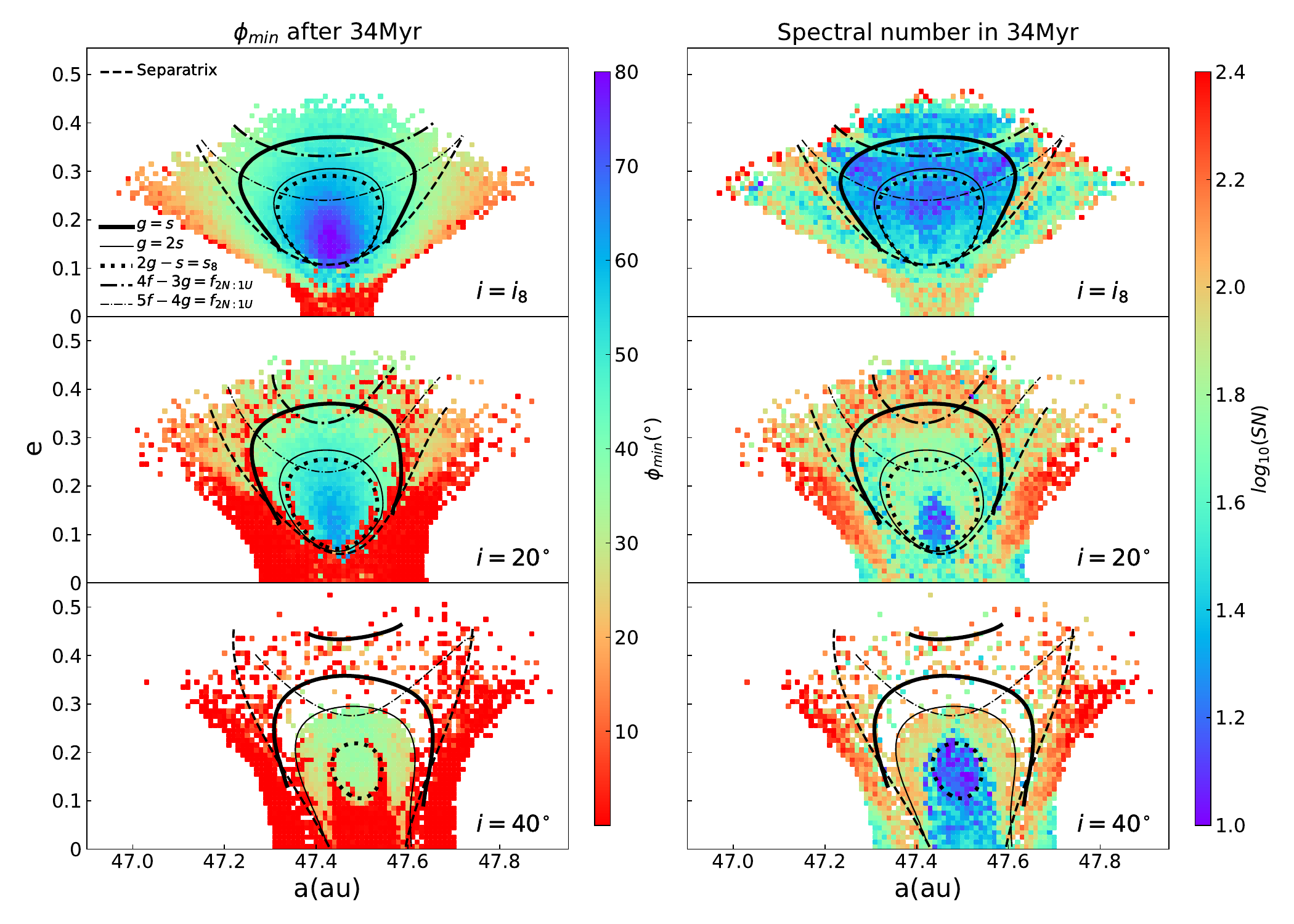}}
\caption{Stability map of the 1:2 resonance in the $(a,e)$ plane. The $\phi_{\rm min}$ and the SN are indicated by colour in the left and the right column, respectively. The locations of the secular mechanisms (see text), and the boundary between horseshoe resonance and asymmetric resonance islands are plotted as lines of different types.}
\label{fig:1-2_ae}
\end{figure*}

Adopt the $\phi_{\rm min}$ and SN as indicators, we present maps of stability on the $(a,e)$ plane in Fig.~\ref{fig:1-2_ae}. Some secular mechanisms as discussed in the $(a,i)$ plane are also plotted. 

Since the proper frequency $s$ is almost exclusively related to the orbital inclination, in the $(a,e)$ planes of $i=i_8 \approx 0^{\circ}$ and $i=20^{\circ}$, the locations of mechanisms $g=2s$ and $2g-s=s_8$ almost coincide with each other. In fact, this is a natural result of the almost constant value of the proper frequency $s$ that satisfies $3s \approx s_8$. We note that the SN increases slightly due to the combined effect of $g=2s$ and $2g-s=s_8$.

The effect of $2g-s=s_8$ on $\phi_{\rm min}$ cannot be easily recognised at $i=i_8$, while at $i=20^{\circ}$ and $i=40^{\circ}$, many particles with $e\lesssim0.2$ have smaller $\phi_{\rm min}$. This is because their eccentricity is reduced to very low value by $2g-s=s_8$, which subsequently triggers the switching between resonance islands, as explained previously. Regardless of the inclination, $2g-s=s_8$ brings little change in the SN, implying that its effect on the orbital stability is very limited.

On the contrary, the effect of the ZLK mechanism increases significantly with increasing inclination. As we can see in Fig.~\ref{fig:1-2_ae}, it slightly increases the SN at $i=i_8$, but significantly excites it at $i=20^{\circ}$. At $i=40^{\circ}$, the ZLK mechanism effectively clears nearly all the test particles in the high eccentricity region, especially those between ZLK mechanism and the $g=2s$. At this inclination, the $g=2s$ acts as a wall delineating the boundary of the stable region.

In addition to the three secular mechanisms mentioned above, at $i=i_8$ we can also find secondary resonances related to the quasi 2:1 MMR between Uranus and Neptune and the proper frequency $f$, including $4f-3g=f_{2N:1U}$ and $5f-4g=f_{2N:1U}$, where $f_{2N:1U}$ is the frequency of $2\lambda_8-\lambda_7$. These resonances may increase the SN and lower the value of $\phi_{\rm min}$, with the effect of $4f-3g=f_{2N:1U}$ being more pronounced due to its lower order.

\subsection{1:2 MMR in $(e,i)$ plane} \label{sec:switch}

\begin{figure}[!htb]
\centering
\resizebox{\hsize}{!}{\includegraphics{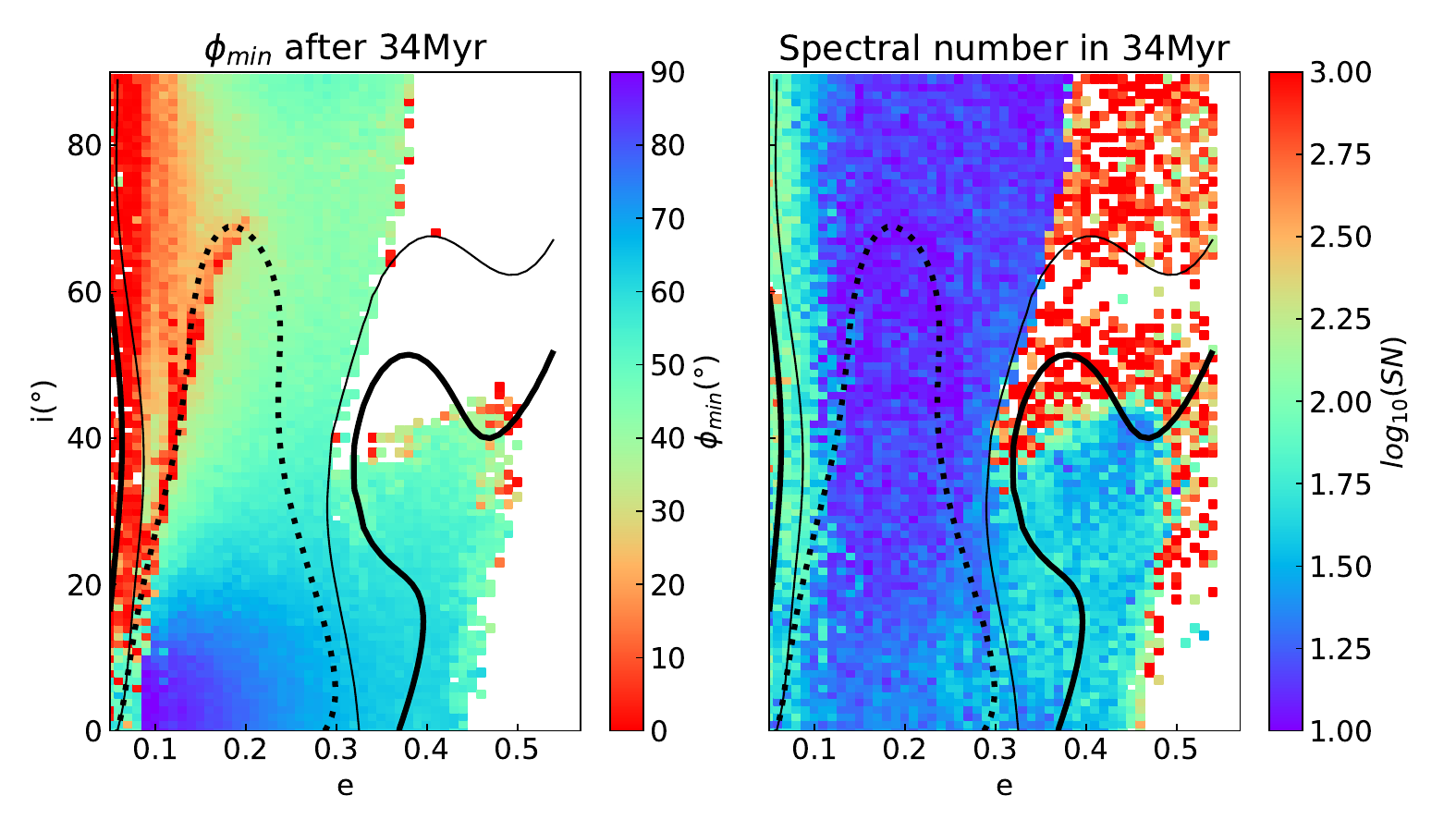}}
\caption{The same as Fig.~\ref{fig:1-2_ae}, but in the $(e,i)$ plane. }
\label{fig:1-2_ei}
\end{figure}

We further explore the 1:2 resonance in the $(e,i)$ plane for a comprehensive understanding of its structure. Fixing $\Omega=\Omega_8$ and $M=M_8$  and regarding $a$ and $\omega$ as the functions of $e$ and $i$, we take the initial eccentricity and inclination from a grid on the $(e,i)$ plane, and the values of $a$ and $\omega$ are calculated through test runs to put the initial orbits in the corresponding centre of the 1:2 MMR.  Then thousands of orbits from these initial conditions are numerically integrated and the stability maps are constructed, as in Fig.~\ref{fig:1-2_ei}.

Fig.~\ref{fig:1-2_ei} corroborates our results presented in previous sections. Since all test particles are placed along the resonance centre, the region of instability on the $(e,i)$ plane is mainly in the area with high eccentricity and high inclination. The ZLK mechanism appears in the high eccentricity region, and it strongly destroys the stability of orbits with high inclination. The combined effects of the ZLK mechanism and the $g=2s$ dominate the high inclination region and forms the most unstable area between these two mechanisms. 

The asymmetric resonance is hardly to retain for long time when the eccentricity is less than 0.1, even if the test particles have been carefully placed at the resonance centre. The secular resonance $2g-s=s_8$ prominently influences the regions with eccentricity smaller than 0.2, and may cause particles to switch between two asymmetric islands and reduce the proper frequency $f$. Its effect becomes less pronounced at higher eccentricities because it influences the motion mainly by causing oscillation of eccentricity. We might expect that $2g-s=s_8$ will play a significant role in altering the population ratio between the leading and trailing asymmetric islands, because the switches between these two asymmetric resonance islands will certainly mix and finally equally divide the population at low eccentricity. We note that the position of small value of $\phi_{min}$ in the left panel of Fig.~\ref{fig:1-2_ei} aligns well with the regions of high `fragility' observed in \citetads[][Figs. 13 \& 26 therein]{Gallardo2020}, suggesting that the fragility in the 1:2 resonance may be associated with the secular resonances discovered in our paper. In addition, the upper left corner (low eccentricity and high inclination) of very small $\phi_{\rm min}$ in Fig.~\ref{fig:1-2_ei} agrees very well with the fact that the observed Twotinos are absent in this region, implying that $\phi_{\rm min}$ is a good indicator of  orbital stability.

\subsection{1:3 MMR}

We applied the similar methods and techniques to the 1:3 MMR to obtain a comparison and a reference for the 1:2 MMR.  To achieve a high resolution in the frequency analyses especially for low frequency domain, we extended both the integration time and output interval by a factor of four, so that the integration time is $\sim$134\,Myr. The set of initial conditions is similar to that for the 1:2 resonance. We briefly summarise the results for the 1:3 MMR in two sets of dynamical maps on the $(a,i)$ and $(a,e)$ planes. As before, the important secular mechanisms have been figured out. 

\begin{figure}[!htb]
\centering
\resizebox{\hsize}{!}{\includegraphics{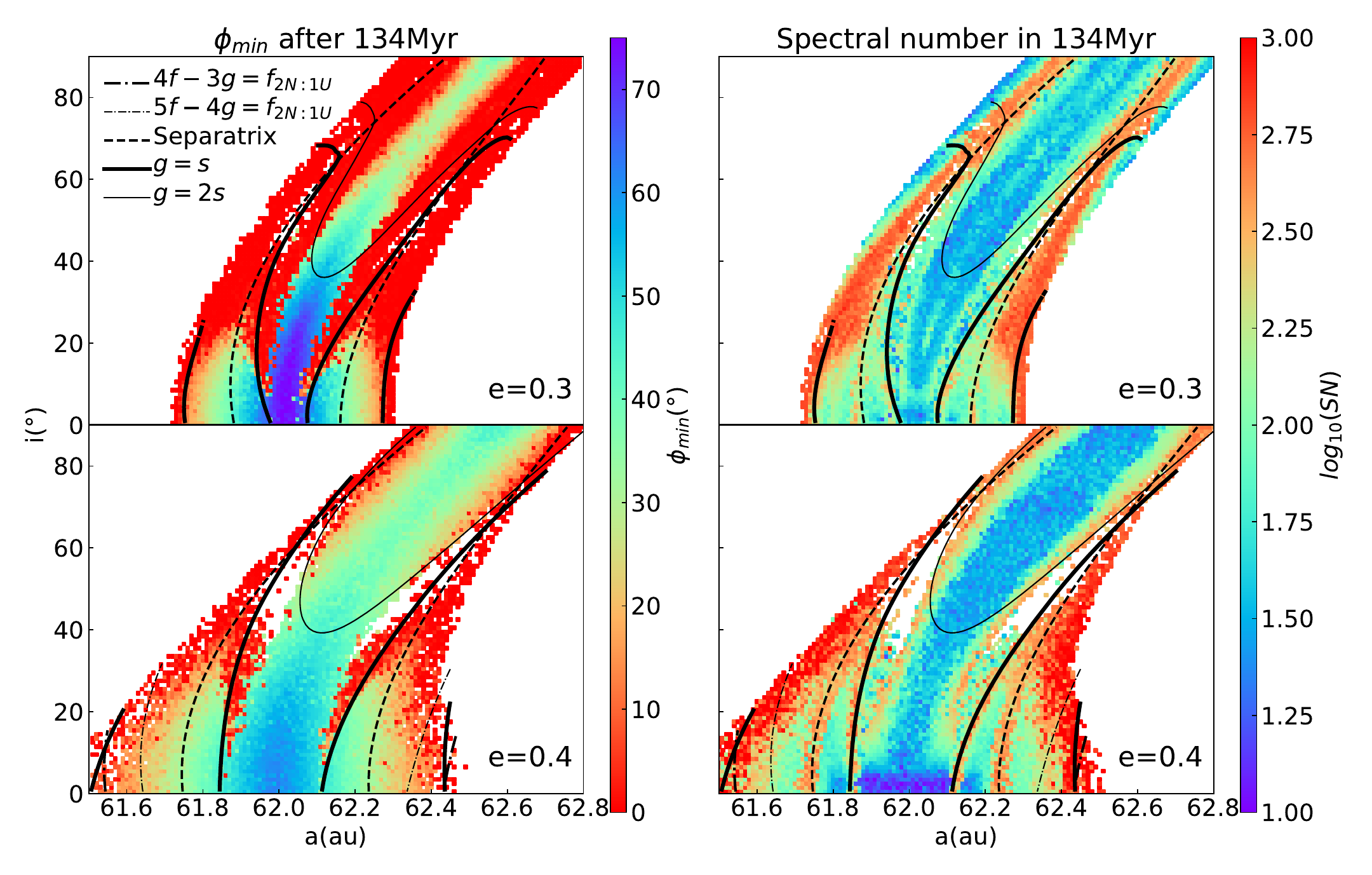}}
\caption{Dynamical maps of the 1:3 resonance on the $(a,i)$ plane. Two indicators, the $\phi_{\rm min}$ (left column) and the SN (right column), have been adopted. Two initial eccentricity values, $e=0.3$ (top panels) and $e=0.4$ (bottom panels), have been used to indicate the stability of orbits. Recognised secular mechanisms and the separatrix between the horseshoe and asymmetric resonances are plotted in lines.}
\label{fig:1-3ai}
\end{figure}

In Fig.~\ref{fig:1-3ai}, we show the dynamical maps on the $(a,i)$ plane. Compared to the 1:2 MMR, the 1:3 MMR maintains a relatively intact structure due to its larger distance from the planet. Compared to the results in the CR3B model \citepads[e.g. Fig.~3 of][for initial eccentricity $e=0.6$]{Gallardo2020}, the resonance island in $(a,i)$ plane has the similar shape (for prograde orbits, as we have only calculated orbits with $i<90^\circ$), but the motion in Fig.~\ref{fig:1-3ai} seems less stable.  In addition to the boundary between horseshoe and asymmetric islands where the SN is relatively high, the most remarkable structure arises from the ZLK mechanism ($g=s$), which can be seen easily in both $\phi_{\rm min}$ and SN. In the low inclination region, the ZLK mechanism appears in the inner part of the asymmetric islands, which differs from the situation in 1:2 MMR. In the region with inclinations larger than $40^\circ$, the ZLK mechanism significantly empties the neighbourhood area. The $g=2s$ mechanism only presents in the high inclination region and is somewhat different from the case in 1:2 resonance. However, similar to the 1:2 MMR, the area where the $g=2s$ and ZLK mechanisms overlap is the most unstable region. 

\begin{figure*}[!htb]
\centering
\resizebox{\hsize}{!}{\includegraphics{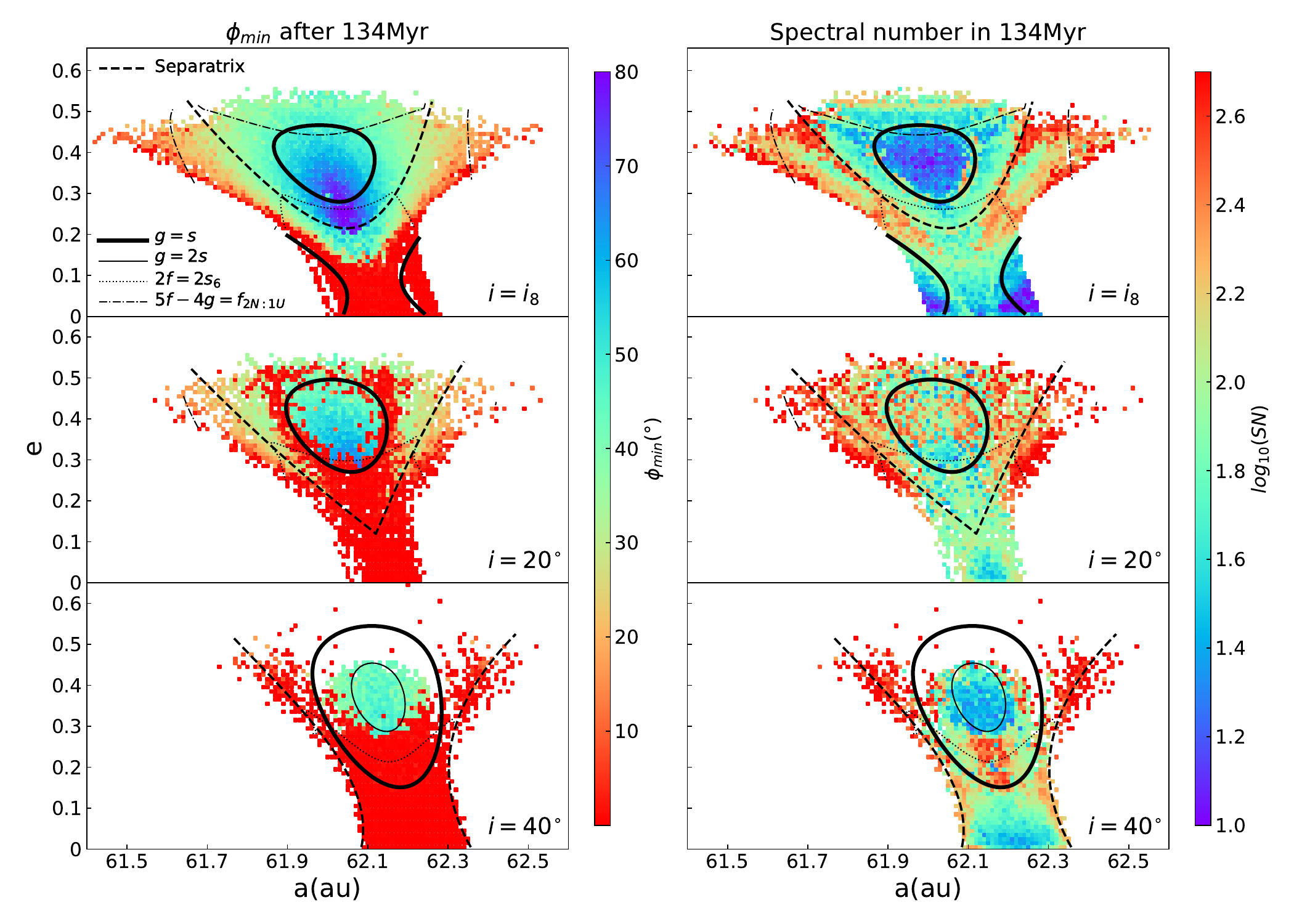}}
\caption{The same as Fig.~\ref{fig:1-2_ae} but for the 1:3 MMR. }
\label{fig:1-3ae}
\end{figure*}

In the 1:3 MMR, the location of ZLK mechanism and the boundary of the asymmetric resonance island no longer coincide, allowing us to better observe the effect of ZLK mechanism. From Fig.~\ref{fig:1-3ai}, it can be seen that the ZLK mechanism extends to lower inclination region, where both $\phi_{\rm min}$ and the SN reveal its influences. Even when the inclination is almost $0^\circ$ ($i=i_8$) in the $(a,e)$ plane (Fig.~\ref{fig:1-3ae}), an increase in SN can be observed, implying it's taking effect. Nevertheless, the ZLK mechanism's influence becomes weaker toward low inclination region, as we can see in Fig.~\ref{fig:1-3ai}. 

In Fig.~\ref{fig:1-3ae} of dynamical maps on the $(a,e)$ plane with given inclinations $i=i_8, 20^\circ, 40^\circ$, we see that the most stable orbits with small SN values can be found at relatively high eccentricity region ($e\sim0.4$), and this is the reason why only two dynamical maps with eccentricity $e=0.3, 0.4$ have been presented in Fig.~\ref{fig:1-3ai}. The small SN value of some orbits at low eccentricity region ($e\lesssim0.1$) seemingly implies that they are stable (regular) orbits. But in fact, with such small eccentricity, these near circular orbits are not locked in the 1:3 MMR, thus not in the scope of this paper. In the CR3B model, the 1:3 resonance exhibits a resonance width close to zero when $e\sim 0$, and there are no asymmetric resonance islands for $e \lesssim 0.1$ \citepads{Lan2019, Gallardo2020}. It is possible that by selecting carefully the appropriate initial parameters we might find some orbits in the 1:3 resonance at such low eccentricity. However, in the outer Solar System model, stable orbits in such low eccentricity region are relatively rare. Since these low eccentricity particles are not in the resonance, their $\phi_{\rm min}$ may approach zero, and this explains the apparent `discrepancy' between the $\phi_{\rm min}$ and SN in the low eccentricity region in Fig.~\ref{fig:1-3ae}. We note that this can also be found in Fig.~\ref{fig:1-2_ae} for the 1:2 MMR. 

The secular resonances associated with the quasi 1:2 MMR between Uranus and Neptune becomes higher in order because the libration period of the resonance angle becomes longer in the 1:3 MMR. Thus, few effect associate with these secular resonances can be seen. However, in the region of $e\lesssim0.3$ where the resonance angle librates particularly slowly, the frequency $f$ approaches $s_6$, resulting in an increase in the SN. It should be noted that the $2f=2s_6$ also exists (but is not shown) in Fig.~\ref{fig:1-3ai} at $e=0.3$, which appears around the resonance separatrix and does not have a significant effect.

In summary, the libration and precessions in the resonances as distant as the 1:3 MMR are generally very slow, and thus the simple integer ratios between these frequencies and those of major planets are rare. As a result, the dynamic map of distant resonances ought to be simpler and cleaner. It can be speculated that in more distant 1:N resonances, the secular mechanisms that play a role in the dynamics of resonant orbits should be limited to mechanisms like the  ZLK and $g=2s$ mechanisms, rather than $2g-s=s_8$ or $4f-3g=f_{2N:1U}$. 

\section{Distribution of Observed Twotinos}

Many studies have attempted to predict the population ratio of small objects (Twotinos) locked in the two asymmetric islands of the 1:2 MMR by simulating planetary migration \citepads[e.g.][]{Chiang2002, MurrayClay2005, Li2014b, Pike2017, Li2023}. However, as we have shown, the secular resonance $2g-s=s_8$ exists widely inside the 1:2 MMR and can drive objects' eccentricities to oscillate. A very low eccentricity can eliminate the asymmetric resonance islands or even cause a temporary break of the 1:2 MMR, resulting in the switching of Twotinos between leading and trailing islands. Because of the existence of such switches between resonance configurations via the low eccentricity channel, it is particularly difficult (if not impossible) to trace back the evolution of the Solar System billions of years ago by simply examining the population ratio between asymmetric resonance islands. Even worse,  observational evidence now tends to suggest that there is no asymmetry in population in the 1:2 MMR \citepads{Chen2019}. 

\citetads{Li2023} show that Neptune’s outward migration can result in different distributions of eccentricities in the two asymmetric islands, specifically particles in the leading island may have higher eccentricities. As we have shown, the secular resonance $2g-s=s_8$ works profoundly in low eccentricity region, mixing the low-eccentricity populations from the two asymmetric islands. More importantly, since the eccentricities of objects in the leading island captured during the planets' migration are higher than the ones in the trailing island \citepads{Li2023}, this switching and mixing of low-eccentricity particles aided by the $2g-s=s_8$ will effectively drive particles from the trailing island to the leading island, and consequently increase the population ratio between these two islands.  

\begin{figure}[!htb]
\centering
\resizebox{\hsize}{!}{\includegraphics{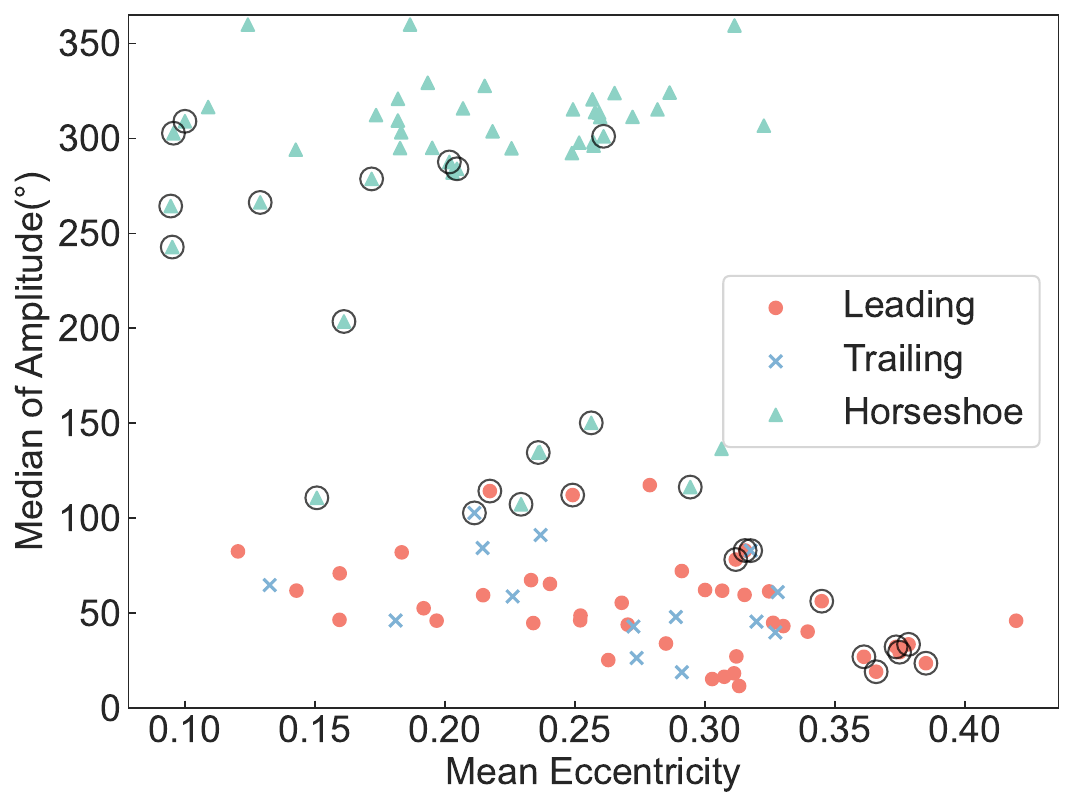}}
\caption{Mean eccentricities and median of libration amplitudes (see text for the definition of median of amplitude) of observed Twotinos. The solid circles, crosses and solid triangles represent the leading, trailing, and horseshoe Twotinos, respectively. Objects that experience the secular resonance $2g-s=s_8$ are circled.}
\label{fig:observation}
\end{figure}

A list of observed Twotinos, now consisting of 107 confirmed objects and 24 candidates, can be found on the website {\it List of Known Trans-Neptunian Objects}\footnote{https://www.johnstonsarchive.net/astro/tnoslist.html}. We downloaded the orbital elements of these objects from the {\it Asteroids-Dynamic Site}\footnote{https://newton.spacedys.com/astdys} (AstDyS) and then determined their orbit configuration by numerically simulating their motion. Among the 131 objects in the list, except for 11 objects that cannot be calculated due to inadequate observations or unavailable data, we finally found 15 objects that are not in the 1:2 MMR, and the rest 105 objects reside in the resonance. Among them, more than 80 objects visit both asymmetric resonance islands in various time during the integration of 34\,Myr. An object is classified as an asymmetric Twotino only if it resides in one of the asymmetric islands and has never undergone an island switching. According to this rigorous criterion, we found 43 Twotinos in the leading island (leading Twotinos), 14 trailing Twotinos, and 48 horseshoe Twotinos. Surely, by this definition, those objects that experience the island switching are classified as `horseshoe Twotinos'. We plotted in Fig.~\ref{fig:observation} the eccentricities and the libration amplitudes of these objects. 

The median of amplitude in Fig.~\ref{fig:observation} is calculated as follows. The total integration time (34\,Myr) is divided into 20 `windows'. We then calculated the libration amplitude of the resonance angle in each window, and the median value was obtained from these 20 amplitudes. The mean eccentricity in Fig.~\ref{fig:observation} is simply the algebraic mean value during the integration. 

In Fig.~\ref{fig:observation}, none of the particles in the asymmetric islands has eccentricity below 0.1. This is consistent with the results in Fig.~\ref{fig:1-2_ei}, where the lower limit of eccentricity required to maintain the asymmetric resonance is approximately 0.09. It is worth noting that due to the perturbations from major planets, the critical eccentricity for maintaining the resonance in the outer Solar System model is higher than the ideal value ($\sim$0.04) in the planar CR3B model \citepads{Malhotra1996, Lan2019}. The highest eccentricity for Twotinos in Fig.~\ref{fig:observation} is approximately 0.42, agreeing with Fig.~\ref{fig:1-2_ae} and Fig.~\ref{fig:1-2_ei}, where the upper limit of eccentricity is approximately 0.45.

The most interesting feature in Fig.~\ref{fig:observation} is that the leading Twotinos have relatively higher eccentricities than the trailing ones, a finding that is consistent with the predictions by \citetads{Li2023}. Specifically, the 10 Twotinos with the highest eccentricities are all located in the leading island, while the highest eccentricity among the trailing Twotinos is about 0.33. \citetads{Chen2019} have shown that the distribution of eccentricity for both asymmetric islands conforms to a Gaussian distribution with a centre of 0.275 and a width of 0.06. However, we obtain from the observational data a difference of 0.0233 between the mean eccentricities of the leading and trailing Twotinos. It does not necessarily mean that the result in this paper contradicts to the findings of \citetads{Chen2019}. In fact, the difference in mean eccentricity found by them (0.0394) between the 17 leading Twotinos and 8 trailing Twotinos is even larger. It is important to note that both our sample of 57 Twotinos and their sample of 25 Twotinos are not large enough to draw very solid conclusion, and to confidently reject the hypothesis that `the asymmetric islands have identical eccentricity distributions'.

The effect of the $2g-s=s_8$ is found to be significant in the motion of Twotinos. In Fig.~\ref{fig:observation}, 28 Twotinos that are found to experience the $2g-s=s_8$ in the numerical integrations have been indicated.  According to their dynamical behaviour, we broadly classify these objects into three groups. The 10 Twotinos with eccentricity above 0.3 belong to the first group. All of these 10 particles, 9 in the leading island and 1 in the trailing island, stay invariably in the respective asymmetric islands without experiencing island switching. This implies a limited effect of the $2g-s=s_8$ in this (relatively) high eccentricity region. 

The second group comprises 15 horseshoe Twotinos affected by the $2g-s=s_8$. A majority of these objects have undergone island switching during the integration. Among them, 5 Twotinos exhibit relatively low median amplitudes ($<160^{\circ}$), indicating that they predominantly experience the asymmetric resonance throughout the simulation, despite having larger amplitudes compared to objects consistently residing within either one of the asymmetric islands. Additionally, there are 10 objects with relatively large amplitudes ($>200^{\circ}$), suggesting a higher likelihood of being always in the horseshoe resonance, i.e. in the horseshoe-like orbits surrounding both the asymmetric islands. 

The third group consists of only three Twotinos. They have relatively low eccentricities ranging from 0.21 to 0.25, and remain in the asymmetric resonance (2 in leading and 1 in trailing island). These objects have median amplitudes $>100^{\circ}$, suggesting that while they maintain themselves in the asymmetric resonance during the 34\,Myr simulation, the island switching is still likely to occur over longer timescales. Empirically, objects with median amplitudes ranging from $100^{\circ}$ to approximately $290^{\circ}$ are predominantly influenced by the $2g-s=s_8$, indicating a strong association between the $2g-s=s_8$ and the occurrence of island switching. 

Therefore, the presence of the $2g-s=s_8$ secular resonance may influence the distribution of eccentricities and the population ratio of Twotinos in the two asymmetric resonance islands of the 1:2 MMR, particularly for those objects with low to medium eccentricities. Twotinos in either resonance modes of the 1:2 MMR in the low eccentricity region ($e\lesssim0.2$) may have been mixed up by the $2g-s=s_8$. Any analysis aiming to reconstruct the planetary migration history using the information of Twotinos in the asymmetric islands of the 1:2 MMR should take into account the influence of the $2g-s=s_8$. Although the number of Twotinos with low eccentricity is still relatively small, it is recommended to exclude those objects that undergo island switching (largely due to $2g-s=s_8$) from the statistics, as they are likely not in the original resonance islands, and any information related to planetary migration within them may have been distorted.

Besides the eccentricity and libration amplitude, we also checked Twotinos' inclinations as well as their evolutions during the integration. We found that the highest inclination of observed Twotinos is $\sim$30$^\circ$, lower than the critical inclination at which the ZLK mechanism is expected to have a significant effect, as shown in Fig.~\ref{fig:sn} and Fig.~\ref{fig:1-2_ei}. This suggests that all the observed Twotinos are in a safe range of inclinations and are not subject to strong  ZLK oscillations, even if their proper frequencies meet the conditions of the ZLK mechanism.

\section{Conclusions}

In the trans-Neptunian region, the 1:N mean motion resonances with Neptune are of particular interest, because in these MMRs the symmetric resonance configuration (with the resonance angles librating around $180^\circ$ or $0^\circ$), the asymmetric configuration and the horseshoe resonance coexist.  And important clues to the early history of the Solar System may be found in the populations of TNOs trapped in these MMRs. In this paper, we conducted systematic analyses on two Neptunian resonances, the 1:2 MMR and 1:3 MMR. 

Our investigation on these two resonances are basically based on the numerical simulations of test particles' motion in the outer Solar System model. To find the representative orbits in the resonances, we chose carefully the initial orbital elements of test particles through test runs to put all the initial conditions in the centre of the resonance (Figs.~\ref{fig:resonance_center}, \ref{fig:prerun}). 

Using the method of frequency analysis, we determined the proper frequencies in the motion of test particles (Fig.~\ref{fig:eigenfre}), including the frequency of the resonance angle's libration ($f$), the precession rates of perihelion ($g$) and of the ascending node ($s$). With these proper frequencies, we then identified the secular mechanisms that may occur and influence the motion in the MMRs. Also by frequency analysis, we obtained the power spectrum of the critical angles, from which the spectral number (SN) was calculated. Adopt the SN as the indicator of regularity (stability) of orbits, we constructed dynamical maps on different representative planes, both for the 1:2 MMR (Figs.~\ref{fig:sn}, \ref{fig:1-2_ae}, \ref{fig:1-2_ei}) and for the 1:3 MMR (Fig.~\ref{fig:1-3ai}, \ref{fig:1-3ae}). The locations of those identified secular mechanisms were found to match the structures in the dynamical maps. The behaviours of some typical orbits confirm directly the effects of these secular mechanisms (Figs.~\ref{fig:example}, \ref{fig:example2}). 

To better distinguish the resonance modes of orbits, we also define the minimum of resonance angle ($\phi_{\rm min}$) in this paper, and it was found to be able to tell the stability of an orbit in some sense  (Fig.~\ref{fig:lifetime}). The lifespan is the most straight measure of the orbital stability, but generally it is very expensive to compute the lifespan. Fortunately, both the $\phi_{\rm min}$ and the SN can be calculated easily from numerical simulations of motion for relatively short time, and they are tightly related to the lifespan (Fig.~\ref{fig:lifesn}).

As have been plotted over the dynamical maps (Figs.~\ref{fig:sn}, \ref{fig:1-2_ae}, \ref{fig:1-2_ei}, \ref{fig:1-3ai}, \ref{fig:1-3ae}), the most significant mechanisms we detected in the 1:2 and 1:3 MMRs include the ZLK mechanism and the $g=2s$ mechanism. Since their critical angles do not satisfy the D'Alembert's rule, they are just referred as `mechanisms' rather than `resonances'. The ZLK mechanism is the main cause of instability and takes place in a large region. The exchange of eccentricity and inclination during the orbital evolution is a common phenomenon and the ZLK mechanism increases the magnitude of this exchange. The increase in eccentricity of small objects during oscillation causes their perihelion to approach Neptune, finally resulting in destabilization of the orbits. The $g=2s$ mechanism behaves like a weakened version of the ZLK mechanism. In our simulations, the region of high inclination between the $g=2s$ and the ZLK mechanisms is the least stable region and would be rapidly evacuated in tens of millions of years. 

The existence of TNOs with high inclination has always been a puzzling issue. The oscillation of the ZLK mechanism inside 1:2 or 1:3 MMR is on the order of ten million years, which makes it possible for TNOs to obtain the high inclination by trading their eccentricities through the ZLK mechanism over millions of years. On the other hand, it is easy for objects locked in the MMRs to obtain high eccentricities as planets migrate outward \citepads[][]{Malhotra1993}. Thus, objects leaking from the 1:2 and 1:3 MMRs might have contributed a considerable fraction of high-inclination population.  

The $2g-s=s_8$ resonance is the only one classical secular resonance that is related with the proper frequency of major planets. It causes a long-period oscillation in the eccentricity and its long-term influence can be observed in the low eccentricity region ($e\lesssim0.2$) of the 1:2 MMR. When the eccentricity drops to a very low level, the asymmetric resonance islands disappear and the protection of the 1:2 MMR is weakened, leading objects to switch between asymmetric islands. Thus, the $2g-s=s_8$ opens the channel for switching between leading and trailing islands. Except this, it brings only little influence on the overall orbital stability, as we can hardly observe any effect of the $2g-s=s_8$ on the SN in Fig.~\ref{fig:sn} and Fig.~\ref{fig:1-2_ei}. Therefore, we expect that it may introduce considerable influences on the population ratio between two asymmetric resonance islands, and on the eccentricity distribution in the relatively low eccentricity region ($e\lesssim 0.2$).

The secondary resonances associated with the quasi 2:1 resonance between Uranus and Neptune are found to present in both the 1:2 and 1:3 MMRs. The relatively strong ones include the $4f-3g=f_{2N:1U}$ and $5f-4g=f_{2N:1U}$ resonances. These resonances tend to appear in the region of high eccentricity and large libration amplitude where the proper frequency $f$ is relatively high. Although not significantly, these secondary resonances introduce more perturbations and thus contribute to the instability of influenced orbits.  

Several previous studies have shown that there are no secular mechanisms other than the ZLK mechanism in the 1:2 resonance \citepads[e.g.][]{Lykawka2007,Tiscareno2009,Li2014b}. This is in rough agreement with our results as we found that the ZLK mechanism is indeed the main reason for destabilizing the orbits of objects in the 1:2 MMR. Other secular mechanisms such as the $g=2s$ and $2g-s=s_8$, and the secondary resonances such as $4f-3g=f_{2N:1U}$ and $5f-4g=f_{2N:1U}$, have been found in this paper, but their effects as revealed by the SN and/or $\phi_{\rm min}$ in the dynamical maps, are weak. 

\citetads{Nesvorny2001} discovered in the 1:2 MMR several secular resonances,  e.g. $2g=2s_8$ at $e\sim0.2$ and $2g-s=s_8$ at $e\sim0.3$. We found in our calculations that although the frequency condition of the former secular resonance ($2g=2s_8$) may be satisfied in a very narrow area at $e\sim0.2, i \sim0^\circ$, the effect of this mechanism can hardly be recognized on the dynamical maps either of $\phi_{\rm min}$ or of SN. As the proper frequencies of Twotinos are often lower than the precession frequencies of major planets, only in the area where $g$ reaches its maximum can it barely match $s_8$. As for the latter secular resonance $2g-s=s_8$, its presence in a wider range ($i$: $0^\circ$--$70^\circ$, $e$: 0.05--0.3) is found in this paper (dotted line in Fig.~\ref{fig:1-2_ei}). It is even more influential in lower eccentricity region when the inclination is relatively higher (see Figs.~\ref{fig:sn}, \ref{fig:lifetime}, \ref{fig:1-2_ae}). We note that the libration of the critical angle of this secular resonance (like Figure 9 in \citetads{Nesvorny2001}) might be easier to be observed at moderate eccentricity ($\sim$0.3) than at lower eccentricity, because in the latter case the complete libration of the critical angle might be destroyed by the asymmetric island switching which is in turn aided by this secular resonance. 

For real Twotinos currently known, the $2g-s=s_8$ is the most influential secular resonance. Out of the 105 observed objects that are confirmed to be in the 1:2 MMR by our calculations, 28 Twotinos are found to be affected by this secular resonance (Fig.~\ref{fig:observation}). It forces the eccentricity of Twotinos to oscillate largely, and in the outer Solar System model it may temporarily weaken the protection from the MMR when the eccentricity is brought to its minimum, where a Twotino may switch from one asymmetric resonance island to the other. Therefore, in the relatively small eccentricity region ($e\lesssim0.2$), Twotinos originally from either asymmetric island have lost their original identity and have been mixed up in the evolution of the Solar System's age.  However, some dynamical properties of the primordial Twotinos may still be preserved in the current population. Combining the results in this paper with the knowledge of planet migration and resonance capture in previous work \citepads[e.g.][]{Li2023}, we can understand very well the eccentricity distribution of current Twotinos, and we expect more evident clues to the Solar System's history to be found in this population, which has been continuously increasing.  

Besides the 1:N resonances, the 2:3 and 2:5 MMRs contain a significant number of TNOs \citepads[e.g.][]{Gladman2012, Nesvorny2015b, Malhotra2018}, but the distribution of TNOs in these non-1:N resonances is less informative since they have only one symmetric resonance island. Other 1:N MMRs than the 1:2 and 1:3 currently might be of less interest because their dynamical structures could probably be relatively simpler due to the great distance from major planets. In addition, observations of these resonances are also much fewer. 

\begin{acknowledgements}
	Our great appreciation goes to the anonymous referee whose insight comments helped us greatly in improving our manuscript.  This work has been supported by the science research grant from the China Manned Space Project with NO.CMS-CSST-2021-B08. We also thank the supports from the National Key R\&D Program of China (2019YFA0706601) and National Natural Science Foundation of China (NSFC, Grants No.12373081, No.12150009 \& No.11933001). 
\end{acknowledgements}

\bibliographystyle{aa-note}
\bibliography{Res1_N}
\end{document}